\documentclass[sigconf]{acmart}
\usepackage{booktabs} % For formal tables
\usepackage{xcolor}
\usepackage{color, colortbl}
\usepackage{multirow,hhline}
\usepackage{pdfpages}
\usepackage{tikz}
\usepackage{fancybox}
\usepackage{array}
\usepackage[usestackEOL]{stackengine}
\usepackage{tikz-qtree}
\usepackage[linguistics]{forest}
\usepackage{xcolor,lipsum}
\usepackage{balance}
\usepackage{stfloats}

\forestset{qtree/.style={for tree={parent anchor=south, 
			child anchor=north,align=center,inner sep=0pt}}}

\usepackage{xargs}  
\usepackage[colorinlistoftodos,prependcaption,textsize=tiny]{todonotes}
\newcommandx{\mynote}[2][1=]{\todo[linecolor=red,backgroundcolor=red!25,bordercolor=red,#1]{#2}}
\tikzset{level distance=60pt,
	sibling distance=6pt,
	every tree node/.style={align=center},
}
\copyrightyear{2018} 
\acmYear{2018} 
\setcopyright{acmcopyright}
\acmConference[ICMI '18]{2018 International Conference on Multimodal Interaction}{October 16--20, 2018}{Boulder, CO, USA}
\acmBooktitle{2018 International Conference on Multimodal Interaction (ICMI '18), October 16--20, 2018, Boulder, CO, USA}
\acmPrice{15.00}
\acmDOI{10.1145/3242969.3243029}
\acmISBN{978-1-4503-5692-3/18/10}

\renewcommand\footnotetextcopyrightpermission[1]{} % removes footnote with conference information in first column
\usepackage{mdwlist}%% compact list formats \itemize*
\begin{document}
	%\title{Interaction Design of a Language Learning Tool for\\Hearing-Impaired Infants Using a Robot
	%and a Virtual Human}
	
	\title{Multimodal Dialogue Management for Multiparty
		Interaction with Infants}

	\author{Setareh Nasihati Gilani}
	\affiliation{
	 \institution{Institute for Creative Technologies}
	\department{University of Southern California}
%	\city{Los Angeles}
%	\state{CA}
%	\country{USA}
	}
	\email{sngilani@ict.usc.edu}
	
	\author{David Traum}
	\affiliation{
	\institution{Institute for Creative Technologies}
	\department{University of Southern California}
%	\city{Los Angeles}
%%	\state{CA}
%	\country{USA}
	}
	\email{traum@ict.usc.edu}
	
	\author{Arcangelo Merla}
	\affiliation{
	\institution{Dept. of Neuroscience \& Imaging Sciences}
	\department{University G. d Annunzio }
%	\state{Chieti}
%	\country{Italy}
	}
	\email{arcangelo.merla@unich.it}
	
	\author{Eugenia Hee}
	\affiliation{
		 \department{Dept. of Computer Science}
			\institution{University of Southern California}
	%	\city{Los Angeles}
	%		\state{CA}
	%		\country{USA}
		}
	\email{hee@usc.edu}
	
	\author{Zoey Walker}
	\affiliation{
     \department{Brain \& Language Lab for Neuroimaging, BL2} 
	\institution{Gallaudet University}
%	\city{Washington}
%	\state{DC}
%	\country{USA}
	}
	\email{zoey.walker@gallaudet.edu}
	
	\author{Barbara Manini}
		\affiliation{
	\department{Brain \& Language Lab for Neuroimaging, BL2} 
	\institution{Gallaudet University}
%	\city{Washington}
%	\state{DC}
%	\country{USA}
	}
	%\email{ barbara.manini@gallaudet.edu}
	\email{b.manini@uea.ac.uk}
	
	\author{Grady Gallagher}
	\affiliation{
	\department{Brain \& Language Lab for Neuroimaging, BL2} 
	\institution{Gallaudet University}
%	\city{Washington}
%	\state{DC}
%	\country{USA}
	}
	\email{grady.gallagher@gallaudet.edu}
	
	\author{Laura-Ann Petitto}
	\affiliation{
		\department{Brain \& Language Lab for Neuroimaging, BL2 } 
	}
	\affiliation{
	\department{PhD in Educational Neuroscience (PEN) Program} 
	\institution{Gallaudet University}
%	\city{Washington}
%	\state{DC}
%	\country{USA}
	}
	
	\email{laura-ann.petitto@gallaudet.edu}
	
%	\end{comment}

	\begin{comment}                % Shared affliation version 2 ( correct but ugly! order is not obvious )
\author{Setareh Nasihati Gilani}
\author{David Traum}
\affiliation{
	\institution{Institute for Creative Technology}
	\department{University of Southern California}
	%	\city{Los Angeles}
	%	\state{CA}
	%	\country{USA}
}
\email{{sngilani,traum}@ict.usc.edu}

\author{Arcangelo Merla}
\affiliation{
	\institution{Dept. of Neuroscience \& Imaging Sciences}
	\department{University G. d Annunzio }
	%	\state{Chieti}
	%	\country{Italy}
}
\email{arcangelo.merla@unich.it}

\author{Eugenia Hee}
\affiliation{
	\department{Dept. of Computer Science}
	\institution{University of Southern California}
	%	\city{Los Angeles}
	%		\state{CA}
	%		\country{USA}
}
\email{hee@usc.edu}

\author{Zoey Walker}
\author{Barbara Manini}
\author{Grady Gallagher}
\author{Laura-Ann Petitto$^*$}
\affiliation{
	\department{PhD in Educational Neuroscience (PEN) Program} 
	\institution{Gallaudet University}
	%	\city{Washington}
	%	\state{DC}
	%	\country{USA}
}
\email{{zoey.walker,barbara.manini,grady.gallagher,laura-ann.petitto}@gallaudet.edu}

	\end{comment}

	\renewcommand{\shortauthors}{S.N. Gilani et al.}

	\begin{abstract}
		We present dialogue management routines for a system to engage in multiparty agent-infant interaction. The ultimate purpose of this research is to help infants learn a visual sign language by engaging them in naturalistic and socially contingent conversations during an early-life critical period for language development (ages 6 to 12 months) as initiated by an artificial agent. As a first step, we focus on creating and maintaining agent-infant engagement that elicits appropriate and socially contingent responses from the baby. Our system includes two agents, a physical robot and an animated virtual human. The system's multimodal perception includes an eye-tracker (measures attention) and a thermal infrared imaging camera (measures patterns of emotional arousal). A dialogue policy is presented that selects individual actions and planned multiparty sequences based on perceptual inputs about the baby's internal changing states of emotional engagement. The present version of the system was evaluated in interaction with 8 babies. All babies demonstrated spontaneous and sustained engagement with the agents for several minutes, with patterns of conversationally relevant and socially contingent behaviors. We further performed a detailed case-study analysis with annotation of all agent and baby behaviors. Results show that the baby's behaviors were generally relevant to agent conversations and contained direct evidence for socially contingent responses by the baby to specific linguistic samples produced by the avatar. This work demonstrates the potential for language learning from agents in very young babies and has especially broad implications regarding the use of artificial agents with babies who have minimal language exposure in early life.

	\end{abstract}
	
	

	 % functional Near Infrared Spectroscopy (fNIRS) neuroimaging, Artificial agents as facilitators of infant language learning, Infant neural sensitivity to language patterns, Critical periods in development, }
	
	\maketitle
	\section{Introduction}
	\label{sec:intro}
	Most dialogue system technology is aimed at enabling natural language dialogues with competent language performers. Even language teaching systems generally strive to introduce a new language to competent users of another language. In this paper we describe the dialogue management approach for a very different population: infant L1 language learners who have yet to achieve language competence. The system, called RAVE (\underline{R}obot, \underline{AV}atar, thermal \underline{E}nhanced language learning tool), focuses on multiparty social interaction and learning elements of a visual sign language, \underline{A}merican \underline{S}ign \underline{L}anguage (ASL) \cite{edwards1979signs}. Two differently embodied agents (a robot and a virtual human avatar) engage in the interaction with the infant (and sometimes a parent). The design of the RAVE interface was described in \cite{scassellati2018teaching}, here we focus on dialogue management to support contingent interaction.

The main long-term goal of RAVE is to develop an augmentative learning aid that can provide linguistic input to facilitate language learning during one widely recognized critical developmental period for language (ages 6-12 months; e.g. \cite{petitto2012perceptual}). Exposure to the patterns of language during this period facilitates infants' acquisition of the phonetic-syllabic segments unique to their native language, vocabulary, syntactic regularities, and ultimately, letter-to-segment mapping vital to early reading and academic success  \cite{petitto2016visual}.  This is particularly important for infants who might not otherwise receive sufficient language exposure. %For example, early caretaker-baby conversational exchange can differ with variation in cultural and/or socioeconomic contexts. 
Of particular concern are deaf babies, many of whom are born to parents who do not know a signed language. Rather than receiving minimal language exposure, this circumstance leaves these deaf babies with no access to an accessible language for well into the second year of life when intensive auditory language training typically begins. To support this long-term goal, our system was designed to engage all infants with early-life minimal language exposure \cite{petittoimpact} (hence, our testing of both hearing and deaf babies described in section \ref{sec:evaluation}). 

As a first step, we ask whether the system can engage infants' attention, and whether the system can elicit contingent conversational behaviors from infants. If so, then we will have identified a novel dialogue system with the potential to facilitate language learning in young infants. In order to accomplish this, the dialogue management routines are designed to attract the baby's attention, infer when it is appropriate to provide linguistic stimuli, and to provide contingent reactions to baby initiatives to maintain the dialogue as a socially contingent interaction. 

Multiple inputs to the system's perceptional component were used, including an eye-tracker (a behavioral measure of attention) and a thermal camera and thermal infrared (IR) imaging (a computational psychophysiological measure of change in ANA/autonomic nervous system response \cite{ioannou2014thermal,teena2014thermal}). The use of thermal IR imaging constitutes one unique design feature. It enabled us to track a baby's changes in emotional engagement when interacting with the agents, as indicated by their ANA responses (e.g., a baby's parasympathetic response indicated prosocial engagement as compared with a sympathetic response associated with disengagement or distress \cite{manini2013mom}). Based on our fNIRS brain imaging of infant neural responses to specific language patterns as concomitant with ANA responses, thermal IR algorithms were created as input to the system to permit the identification of when infants were ``ready to learn'' even before they had the capacity to produce language. This was crucial to our dialogue management system, as it provided the central triggering mechanism as to when the agent should start or cease a socially contingent conversational turn. 

%The remainder of this paper is organized as follows. In section \ref{sec:background} we describe the motivation for this kind of system and our specific goals for the contingent dialogue interaction. In  section \ref{sec:systemdesign} we briefly introduce the system components and logical and communication architecture. In section \ref{sec:interactiondesgin}, we focus on the interaction policy and input and output signals. In section \ref{sec:evaluation}, we describe an evaluation of the system which has been tested in interaction with 8 babies as well as providing general observations and a detailed case study. We conclude the paper with section \ref{sec:conclusion} and discuss possible future work in this area. 
%	\vspace{-0.5cm}
	\section{Background, Motivation and Goals} 
	\label{sec:background}
	%Language is the principal system of expression and communication and is arguably the most prominent cultural tool that sets humans apart from the other species. 
Acquiring language starts from birth through a concert of factors including the maturation of the neural systems that support language processing, observation, and engagement in social interactions \cite{finn2010sensitive}. %Language exposure plays an important role in children's early development of language abilities. 
By around ages 6-10 months, hearing babies learn the finite set of sound phonetic units and phonemic categories of their native languages \citep{kuhl2004early}. Children who do not have sufficient language exposure during this critical period are at risk for delays in cognitive, linguistic, and social skills that can span life \cite{saffran2001acquisition,petitto2016visual}. Interestingly, deaf babies with exposure to visual sign language follow a similar pattern of phonological development in sign language, even though the units are silent and produced on the hands \cite{petitto2004baby,stone2018visual}, including a manual homologue to vocal babbling \cite{petitto1991babbling,petitto2001language}.

However, \citet{higgins1980outsiders} reports that 91.7\% of the deaf individuals %in the \underline{N}ational \underline{C}ensus of the \underline{D}eaf \underline{P}opulation (NCDP) 
come from families where both parents are identified as hearing \cite{schein1974deaf}, where learning a new signed language quickly %(and very quickly to be used fluently with their infant) 
can be challenging. Interventions such as cochlear implants \cite{house1976cochlear} exist that can allow %this population 
some access to spoken language \cite{wilson1991better}, but most of them cannot be deployed until the ages of 18-24 months, which is past an early critical %precise developmental 
period for learning basic phonological units. %While efforts have begun to implant children at younger ages (from $\sim$8 months), precise adjustments and tuning of the device, as well as speech training, still typically begins after ages 18-24 months \cite{nicholas2007will}. 
Thus, many deaf infants might be among this at-risk population due to insufficient language input in early life.
Our work aims to provide visual language input to infants in the critical age period for phonological morphological development. 

%Unfortunately, the simplest solution of just video-recording and playing back visual language is unlikely to work \cite{krcmar2007can}. Even well-designed educational material presented to infants on a screen have resulted in minimal learning gain \cite{richert2011media,krcmar2011word}. \citet{Kuhl9096} found that exposing American infants to  Mandarin Chinese speakers reversed a decline in perception of Mandarin phonetic segments, but exposure only to audio or audio-visual recorded stimuli did not. A physical robot may evoke more interest from this young age than a recording or virtual character on a screen and evoke social responses \citep{meltzoff2010social,arita2005can}, but current robots do not have the facial and hand articulation capability or fluidity of motion needed to convey fluent sign language \cite{kose2012evaluation,toshibaRobot,leyzberg2012physical}. Even robots that have been designed specifically to act as tutors for signed languages \cite{kose2014socially,uluer2015new,aran2009signtutor,kose2011humanoid} often cannot support full range of manual dexterity, fluidity and rhythmic temporal patterning of movement, and facial expressiveness required. 

% In ASL, crucial grammatical information is communicated via the systematic and rhythmic temporal patterning that binds sign phonetic-syllabic units into signs, signs into sign phrases and clauses, and signed sentences \cite{petitto2016visual}, as well as subtle facial cues and body shifting \citep{reilly1990acquisition}

In particular, \citet{petitto2004baby} found that specific rhythmic temporal frequencies of language are important. To capture the attention of infants, we need to provide a linguistic stimulus that matches the rhythmic temporal patterning and facial expressiveness of a natural sign language, such as ASL.  In ASL, crucial grammatical information is communicated via systematic (rule-governed) patterned changes in handshape and specific grammatical modulations of space and movement. The rhythmic temporal patterning binds sign phonetic-syllabic segments into signs, signs into sign phrases and clauses, and signed sentences \cite{petitto2016visual}, as well as grammatical facial expressions and body shifting \citep{reilly1990acquisition}. The use of a virtual character on a screen gives the benefit of having an expressive agent that has the manual dexterity and obligatory facial expressiveness %(both in facial features and posture)
to produce sign language samples as linguistic input \cite{kipp2011sign,schnepp2013generating,pezeshkpour1999development,van2007development,jaballah2013review}. 

	\section{System and Architecture}
	\label{sec:systemdesign}
	%System Desgin 
We adopted a complex multi-party design, involving multiple heterogeneous agents, linguistic stimuli tailored to the target population, and several types of sensory inputs. The system includes two agents (a physical robot and a virtual human) that can provide visual behaviors, as well as several sensor devices for perceptual input: an eye-tracker, thermal camera, and a Kinect. %Contingent interaction between the robot and the virtual human establish them as socially-interacting, conversational agents rather than objects.
%The system was built incrementally over the course of 2 years with different components and protocols.
%In the final design, eye-tracker, thermal-imaging module and behavioral GUI were used as conceptual componentssc. 
%In the final design,

Our hypothesis was that to capture the attention of infants, we should provide a linguistic stimulus that matched the rhythmic temporal patterning found in all natural languages, including natural signed languages, such as ASL. 
%To capture the attention of infants, a unique hypothesis tested here was that this would be optimally achieved if we provided a linguistic stimulus that matched the rhythmic temporal patterning found in all natural languages, including natural signed languages, such as ASL.  
\citet{petitto2004baby} found that babies are sensitive to specific rhythmic temporal frequencies of language in early life. Specific rhythmic temporal patterning binds phonetic-syllabic segments into words, phrases, and clauses in spoken languages, with identical processes occurring in signed languages, whereupon specific rhythmic temporal patterning binds sign-phonetic units into signs, signs into sign phrases and clauses, and signed sentences \cite{petitto2016visual}. Grammatical information is also communicated in ASL via systematic (rule-governed) patterned changes in handshape, eye gaze, grammatical modulations of space and movement, and grammatical body shifting and crucially, facial expressions \citep{reilly1990acquisition,kipp2011sign,schnepp2013generating,pezeshkpour1999development,van2007development,jaballah2013review}. Thus, our use of a virtual character on a screen was a key design feature that provides the benefit of having an expressive agent that produces the optimal rhythmic temporal patterning vital to acquire the phonological building blocks of language, manual dexterity, and obligatory facial expressiveness to produce signed language samples as linguistic input. 

We use a robot since it is a physically-embodied agent, which provided a mechanism to engage the baby, a locus for facilitating attention to the virtual human, and a means to introduce a more natural social conversational setting whereupon agents and baby can occupy varying conversational roles. It has been found that a physical robot can evoke interest and social responses from young babies \citep{meltzoff2010social,arita2005can}, but even robots that have been designed specifically to act as signed language tutors \cite{kose2014socially,uluer2015new,aran2009signtutor,kose2011humanoid}, often cannot support the full range of manual dexterity, fluidity, rhythmic temporal language patterning, and facial expressiveness required \cite{kose2012evaluation,toshibaRobot,leyzberg2012physical}. Hence our reasoning to use a physical robot to act as an initial target for infant attention. We also predicted that contingent interactions between the robot and the virtual human helps establish each agent as socially-interacting conversational partners rather than objects. Our design to use a combined robot+avatar was further supported by previous studies that found the exclusive use of video-recording and playing back visual language is unlikely to work \cite{krcmar2007can,richert2011media,krcmar2011word}. \citet{Kuhl9096} found that exposing American infants to human speakers of Mandarin Chinese reversed a decline in perception of Mandarin phonetic segments, but exposure only to audio or to audio-visual recorded stimuli did not.

\begin{figure}[t]
	\centering
	\includegraphics[width=3.1in]{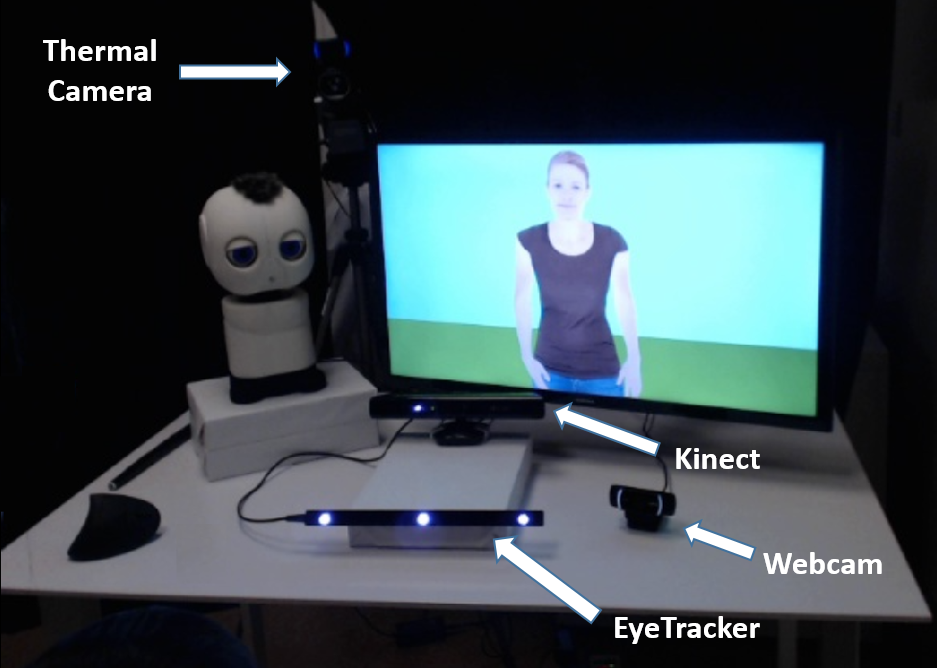}
	\caption{Physical deployment of system components from the front view}
	\label{fig:components}
\end{figure}

Figure  \ref{fig:components} shows the physical deployment of the hardware components from the front view. Multiple webcams were used to record the experiment from different angles. A snapshot of the experiment is shown in Figure \ref{fig:overview}. As seen in Figure \ref{fig:overview}, the infant was sitting on a parent's lap, facing the avatar's monitor and the robot. %The thermal camera, eye-tracker, and Kinect were used as sensors for perceptual inputs. 
In this section we give a brief description about each component and then advance to the architecture of the system. 

\subsection{Components}
\label{sec:components}
%Figure \ref{fig:components} shows a sketch of the physical arrangement of the components in the experimental setup. ({\color{red}TODO: redraw })
A detailed description of each of the components is beyond the scope of this paper. We focus on a brief description to highlight their role in the multiparty dialogue and provide more details on the dialogue manager in
%about how it uses perceptual inputs and the two agents to accomplish this are presented in 
Section~\ref{sec:interactiondesgin}.

\subsubsection{\textbf{Avatar}}
%To capture the attention of infants, we provided linguistic stimuli that matched the rhythmic temporal patterning of language \cite{petitto2016visual} and facial expressiveness of ASL. % In ASL, crucial grammatical information is communicated via the systematic and rhythmic temporal patterning that binds sign phonetic-syllabic units into signs, signs into sign phrases and clauses, and signed sentences \cite{petitto2016visual}, as well as subtle facial cues and body shifting \citep{reilly1990acquisition}. 
The avatar was constructed by capturing a native ASL signer inside of a photogrammetry cage. Facial scans were also captured using a Light Stage \cite{debevec2012light} and the 3-D avatar was built using a real-time character animation system described in \cite{shapiro2011building}. %Figure \ref{fig:avatar} shows the avatar in its idle form. 

\begin{figure}
	\centering
	\includegraphics[width=3.1in]{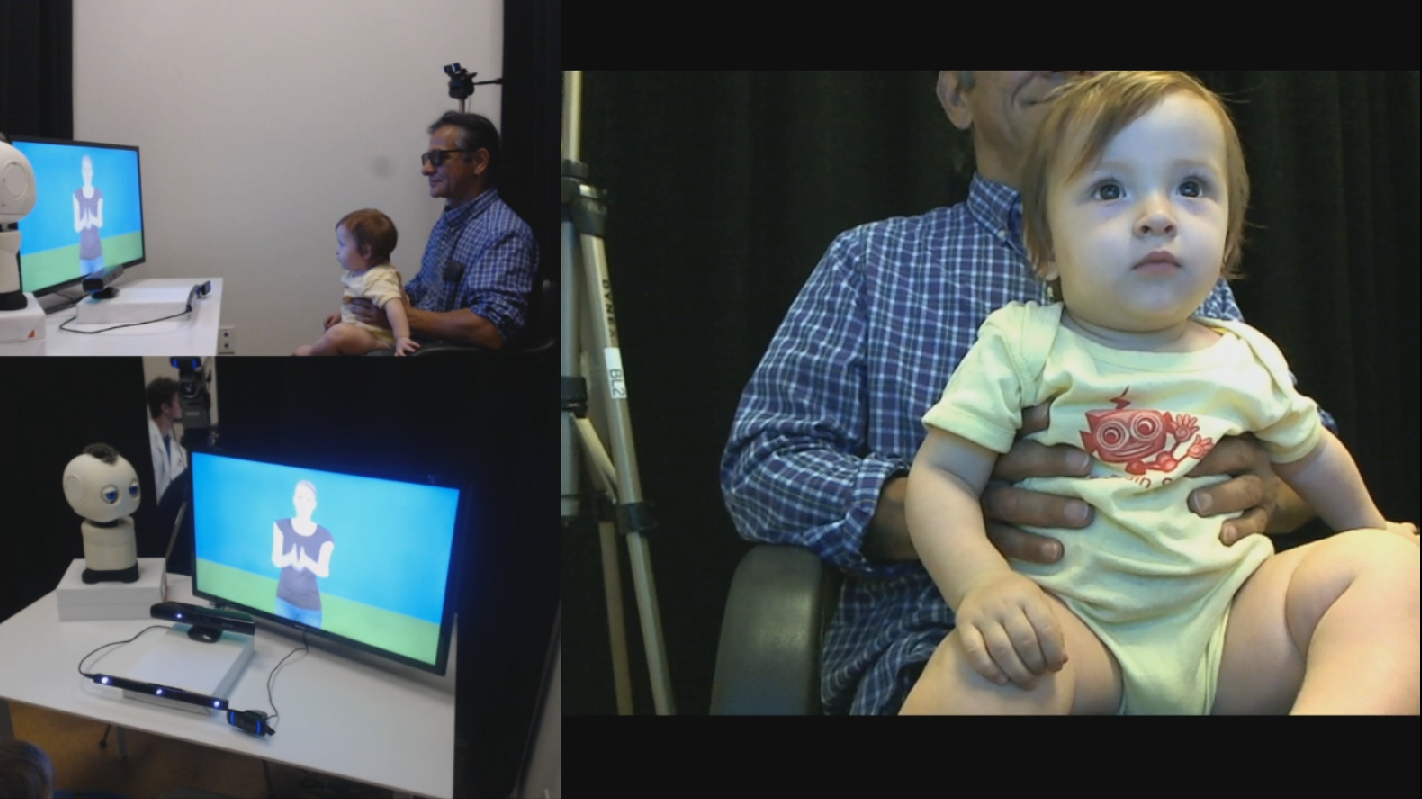}
	\caption{Multiparty interaction between Avatar, Robot, and infant from multiple viewpoints}
	\label{fig:overview}
\end{figure}

% \begin{figure}[h]
% 	\centering
% 	\includegraphics[width=3in]{Figures/3davatar}
% 	\caption{3-D Avatar}
% 	\label{fig:avatar}
% \end{figure}

\subsubsection{\textbf{Robot}}
\begin{comment} % submitted version 
The robot is based on the open-source Maki platform from Hello Robot %\footnote{"https://www.kickstarter.com/projects/391398742/maki-a-3d-printable-humanoid-robot''}.
\cite{makiProject}.
 %We performed some modifications like enlarging the eyes and adding hair to make Maki more baby-friendly according to our specific experimental needs.
 The main purpose of the Robot is to shift the baby's directional attention to the Avatar since it has been shown that infants are capable of following gaze direction \cite{meltzoff2010social}. The robot has an articulated head (pan left/right, tilt up/down), articulated eyes (pan left/right, tilt up/down) and eyelids (open/close). 
\end{comment}
The robot is based on the open-source Maki platform from Hello Robot \cite{makiProject}. 
%The Robot facilitated engaging and shifting the baby's attention to the Avatar and served as one of the socially interactional conversational partners. 
The robot has an articulated head (pan left/right, tilt up/down), articulated eyes (pan left/right, tilt up/down) and eyelids (open/close). \citet{scassellati2018teaching} present more detail about the robot design.

%\subsubsection{\textbf{Dialogue Manager}}

\subsubsection{\textbf{Eye-tracker}}
\begin{figure}[h]
	\centering
	\includegraphics[width=2.9in]{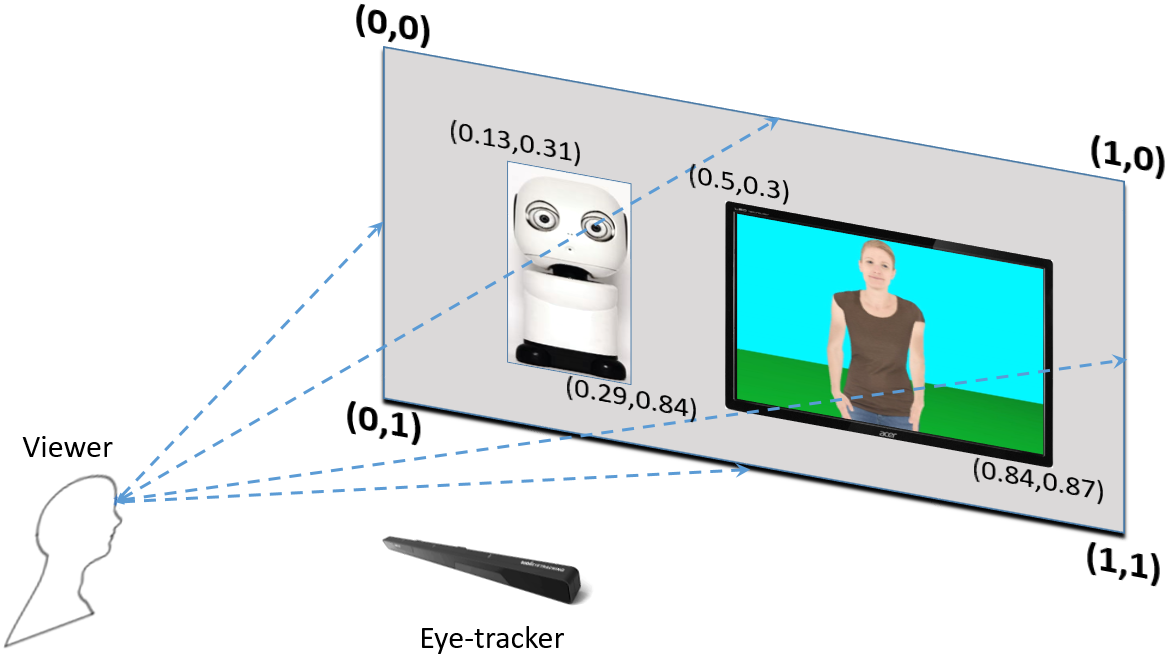}
	\caption{AOI regions from infant's perspective.}
	\label{fig:AOI}
\end{figure}

Infant's gaze direction as a behavioral response was used as an input to the system. A Tobii Pro X3-120 \cite{tobiiEyetracker2} was used to capture the baby's eye-gaze at the rate of 120 Hz using a customized python script. 4 different area of interests (AOI) were defined: Robot, Avatar, In-Between and Outside. AOI coordinates were defined in relation to the infant's point of view, as shown in Figure \ref{fig:AOI}.
We took into account the AOIs as well as the fixation on the target as an indicator of baby's focus of interest. We performed a majority vote paradigm every half second (60 data points) to determine the area of interest. A calibration process was done in the beginning of the experiment to accommodate the program to the physical setup and the relative coordination of the baby's eyes, targets and the tracker. 

\subsubsection{\textbf{Thermal Camera and Thermal Infrared Imaging}}

We used facial thermal patterns and dynamics for capturing the infant's internal state which is used to determine when the infant is engaged with the interaction \cite{ioannou2014thermal}. Thermal infrared imaging, by harnessing the body's naturally emitted thermal irradiation, enables cutaneous temperature recordings to be measured noninvasively, ecologically, and contact free \cite{merla2014thermal}. To calculate information about the infant's affective state, we used the nose tip's average temperature as it was extracted from each frame thus obtaining a temperature signal in real time. The classification of the infant's internal state was built on foundational studies linking the modulation of nose tip temperature and human psychophysiological states, with whereupon positive emotional responses such as interest and engagement are associated with nasal temperature increases (or parasympathetic response) while a decrease in temperature corresponds to sympathetic responses such as distress and disengagement \cite{ioannou2014thermal,manini2013mom}. Thermal IR imaging was performed by means of a digital IR thermal camera FLIR A655sc \cite{thermalImagingCamera} (640 x 480 microbolometer FPA, NETD: < 30 mK @ 30 $^{\circ}$C, sampling rate: 50 Hz).

 \subsubsection{\textbf{Baby Behavior}}
\label{sec:kinect}
\begin{figure}[h]
	\centering
	\includegraphics[width=3.1in]{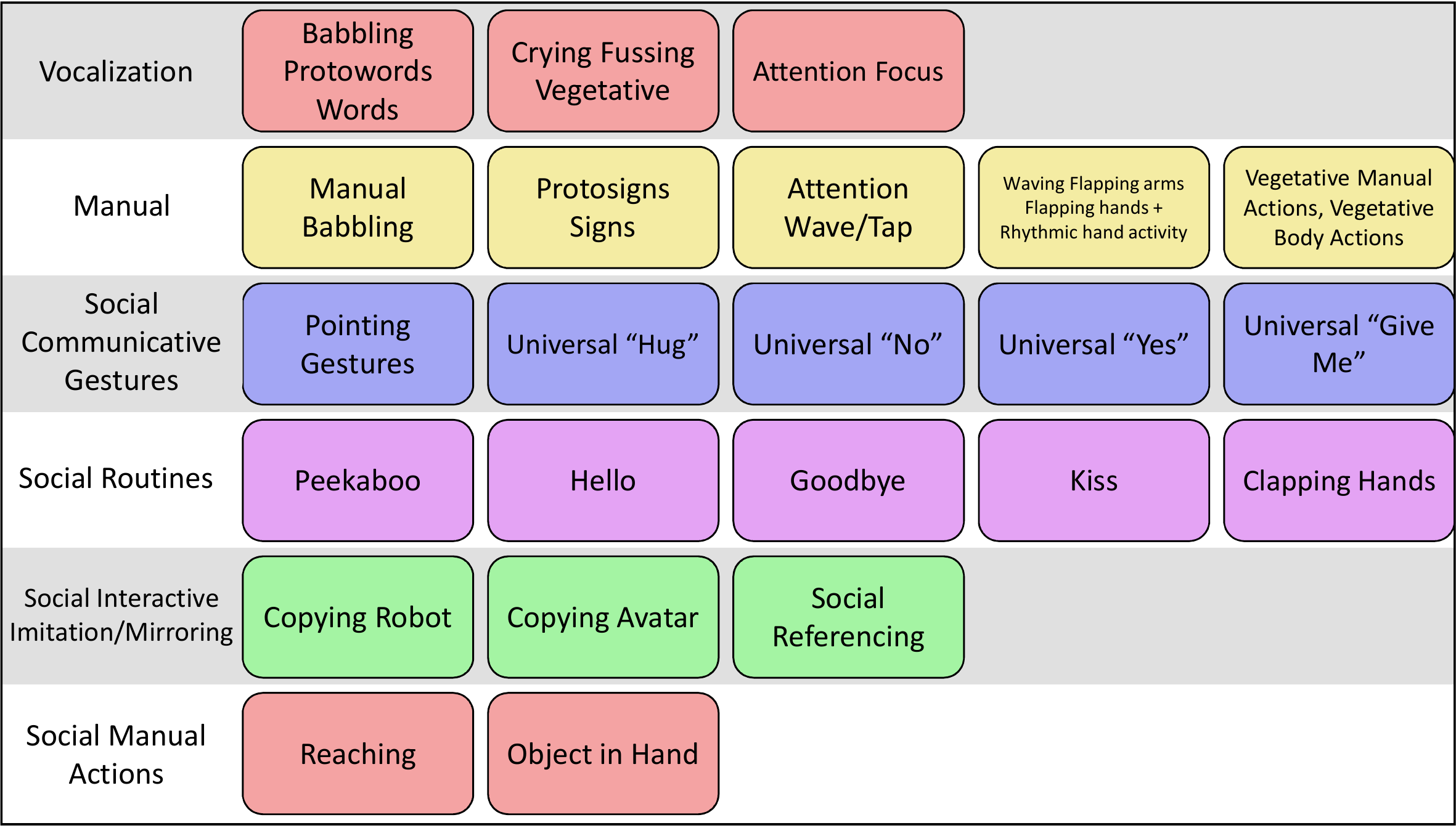}
	\caption{Observer interface for baby's behaviors}
	\label{fig:babyWoz}
\end{figure}

It is crucial to have a visual perception component to capture the communicative and social behaviors of the infant such as hand clapping, pointing, reaching, etc. Current tracking systems such as Kinect \cite{zhang2012microsoft}, use models that are trained on adult anatomy and do not work properly on infants due to their fundamental differences in posture and relative proportions of body parts. Furthermore, the baby is sitting on a parent's lap, so this will bring additional complications for tracking systems which are mostly trained on full body postures. 
To address these issues, we collect Kinect data for future analysis, with the hope of eventually collecting enough data for future customizing models to conform to our specific needs with respect to the experiment setup.
%
%\subsubsection{\textbf{Baby Behavior GUI}}
\label{sec:baby-behavior}
%In the meantime, and until we have enough data for the automatic recognition of baby behaviors, 
But as a first step, we adopted an interface, shown  Figure \ref{fig:babyWoz},  that is used by an expert observer to indicate relevant infant behaviors to the rest of the system in real time .

%\begin{figure}[h]
%\centering
%\includegraphics[width=3in]{Figures/inputsoutputs}
%\caption{Components of the System}
%\label{fig:agents}
%\end{figure}

%- insert diagrams of input/output \\
%- emphasize on the social contingency \\

\subsection{Software Architecture}
\begin{figure}[t]
	\centering
	\includegraphics[width=3.1in]{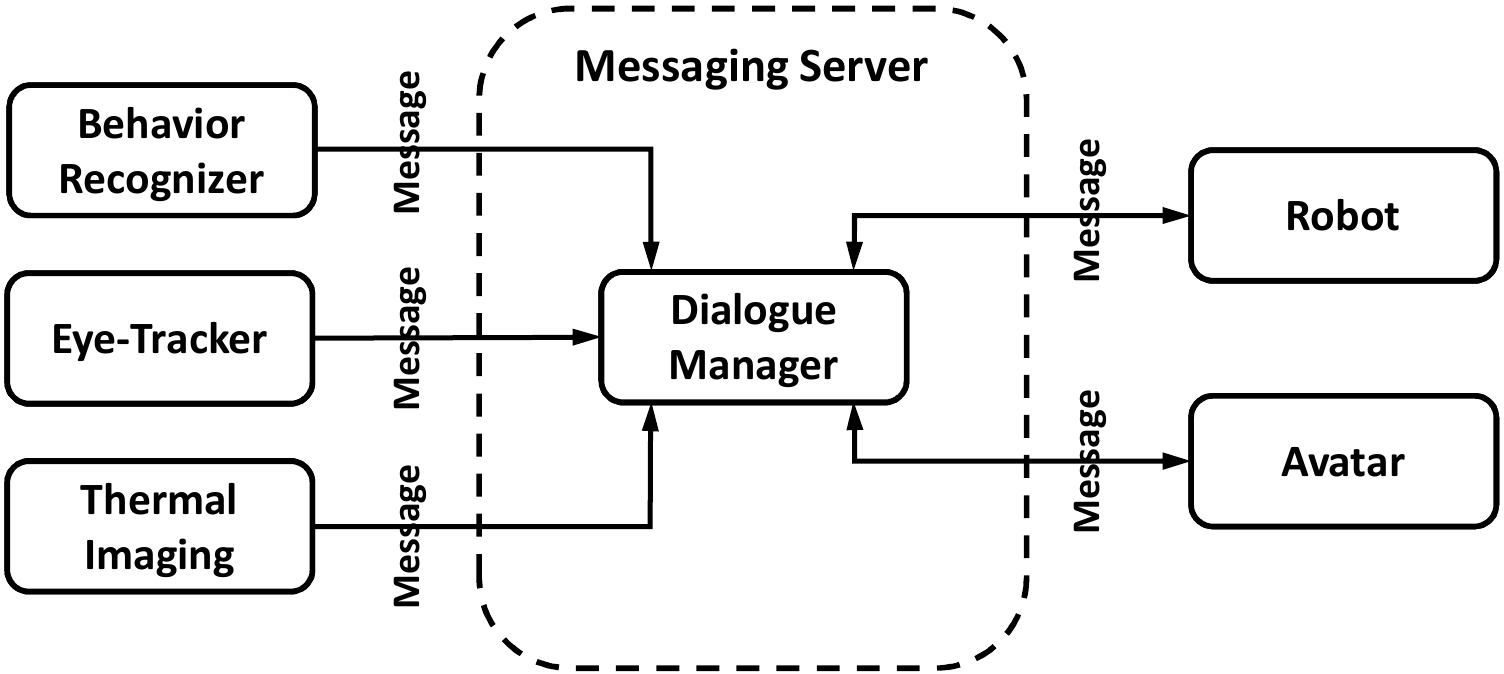}
	\caption{Logical overview of system components}
	\label{fig:architecture}
\end{figure}

Controllers for the hardware components were running on three separate machines, using multiple programming languages. We used a publisher-subscriber model \cite{rajkumar1995real} to facilitate communication between the components where ActiveMQ \cite{activeMQ} was used as a message passing server. As shown in Figure \ref{fig:architecture}, the perceptual components (eye-tracker, thermal imaging and behavior recognizer) send their messages to the server (publishers). These messages are subscribed to by the dialogue controller to update the information state and send messages to the Robot and Avatar (subscribers), directing them to perform communicative behaviors. 

	\section{Dialogue Management}
	\label{sec:interactiondesgin}
	The dialogue manger has three main goals:% (1) To engage the baby by participating in interactive dialogues. (2) To maintain the engagement (sustained engagement). (3)To promote engagement-rendered responses from the baby. 
\begin{enumerate}
	\item To engage the baby by participating in interactive dialogues.
	\item To maintain the engagement (sustained engagement).
	\item To promote engagement-rendered responses from the baby. 
\end{enumerate}

It uses input signals from the perception modules to update its information state \cite{TraumLarsson03} and choose new actions. 
Here, we give more details on multimodal perception signals, different output behaviors for agents, and finally the protocol for deciding on actions.
%system follows a finite state machine architecture where the next action is determined based on input signals as well as the current state following a set of predefined rules. More explanation about these topics will be given below. 

\subsection{Input signals}
\begin{table*}[h]
	\centering
	%\begin{tabular}{|l|l|p{10cm}|}
	{\renewcommand{\arraystretch}{1.5}
		\begin{tabular}{|m{0.33in}|m{1.4in}|m{11cm}| }

			\rowcolor[HTML]{C0C0C0} \hline
			Agent & Category & Behaviors \\ \hline
			\multirow{4}{0.35in}{\centering Avatar} & Conversational Fillers & \ovalbox{Nod}  \ovalbox{Gaze forward/right/left} \ovalbox{Head Shake} \ovalbox{Contemplate}  \ovalbox{Think} \ovalbox{Toss}  \\ \cline{2-3}% \hline
			& Social Behaviors & \ovalbox{Wave}  \ovalbox{Hello}  \ovalbox{Peekaboo} \ovalbox{Go Away / Come Back} \\ \cline{2-3}% \hline %\hhline{~|--|} 
			& Question Solicitation & \ovalbox{What?} \ovalbox{What's Wrong?} \ovalbox{What's That?} \ovalbox{Ready? (To Robot/Baby)} \\ \cline{2-3}% \hhline{~|--|} 
			& Linguistic Patterns & \ovalbox{Good Morning} \ovalbox{Look at Me! (To Robot/Baby)} \ovalbox{Boat} \ovalbox{Pig} \ovalbox{Fish} \ovalbox{Cat}\\ \hline
			\centering Robot & Fillers and Social Behaviors & \ovalbox{Nod} \ovalbox{Hide/Unhide} \ovalbox{Gaze Forward/Right/Left} \ovalbox{Startle} \ovalbox{Blink} \ovalbox{Sleep} \ovalbox{Wake Up} \\ \hline

	\end{tabular}}
	\caption{Robot and Avatar Primitive Behaviors}
	\label{AgentsBehaviors}
	
\end{table*}

These are input signals received from perceptual components and agents as well as the internal dialogue manager signals. Note that in our design, we have no direct perceptual monitoring of the parent, and no Avatar/Robot actions are contingent directly on the parent.

\begin{itemize}
	\item \textbf{Area of Interest (AOI)} is the signal received from the eye-tracker component with discrete values for 4 different areas of the baby's eye gaze: Robot, Avatar, Between and Outside.
	\item \textbf{Readiness-To-Learn} is the signal received from the Thermal Imaging system with 5 discrete values: very negative (sustained decrease in attention, sympathetic), negative (non-sustained decrease in attention, sympathetic), very positive (sustained increase in attention, parasympathetic), positive (non-sustained increase in attention, parasympathetic), and a None signal which shows the signal's absence because of not detecting a reliable signal from the baby.
	\item \textbf{Baby-Behavior (BB)} is the signal received from the human observer interface %which contains information 
about the baby's social and communicative behaviors.
% such as waving, pointing, reaching, etc. 
As seen in Figure \ref{fig:babyWoz}
% shows the interface used to send these commands to the main system. As you can see,
 the input signals are classified into several categories such as vocalization, social communicative gestures, social routines and social manual actions. There are a total of 23 distinct states for this variable. 
	\item \textbf{Component State Signals} come from the Avatar and Robot indicating their state of action: when a requested behavior has started, ended or if there were any errors or exceptions during the execution of a specific behavior.	
	\item \textbf{Timing Signals} are initiated from the dialogue manager itself. The DM tracks when events of different types have happened and sets up automatic signals that can change behaviors, e.g. if nothing interesting has happened recently.
\end{itemize}

\subsection{Output commands}
\label{sec:outputcommands}
There are two different control levels of actions for the agents, described below: Primitive Behaviors and Action Sequences. 
%We provide definitions and examples for each level as follows: 

\subsubsection{\textbf{Primitive Behaviors}}

These are defined as atomic actions of the agents which are single behaviors that cannot be interrupted; such as nodding or a single nursery rhyme. Table \ref{AgentsBehaviors} shows a list of primitive behaviors for the Avatar and Robot. 
%The Robot faces physical limitations due to small number of degrees of freedom in its design, but it is capable of doing basic movements such as turning his head, nodding, blinking and startling with the aim of establishing a social connection with the infant and gain the its attention.

The virtual human's different language samples comprised different conversational/communicative social functions that are commonly used in Infant-Adult conversations. These functions include nursery rhymes, social routines, questions, conversational fillers, soothing responses, social affirmations and negation, solicitations and conversationally neutral idling. These were grouped as follows:
%The current version of the Avatar is capable of doing limited set of behaviors and gestures with respect to 3 categories: 

\begin{enumerate}
	\item \textbf{Conversational Fillers and Social behaviors} Conversational Fillers are short lexical items or phrases that assure the addressee that the conversational partner is attending and still ''in" the conversation. They are like social punctuations e.g., YES! or THAT!, which are full lexical items in ASL. Social behaviors (or, social routines) are standard gestures that are widely used with infants, such as PEEKABOO. %, Negation Head No=full lexical NO in ASL (think analogously, “yeah,” “right,” “go on”)% are mostly used to establish a social connection with the baby and try to make the infant engage in the interaction. The fillers are used to still be active in idle poses. An example can be the think position or slight body movements. {(\color{red} Should I divide this to two categories?)}
	\item  \textbf{Question Solicitation } such as ASL signs WHAT? or WHAT'S THAT? are %particularly 
used when the infant is in a sympathetic 
%or a vegetative 
state. 
	\item \textbf{Linguistic Patterns} provide the vital linguistic stimuli for the baby. All Nursery Rhymes were constructed with the identical rhythmic temporal patterning that matched the infant brain's specific neural sensitivity to that rhythmic temporal patterning \cite{petitto2004baby,petitto2016visual}. The identical overall rhythmic temporal patterning that all Nursery Rhymes were built with was this: maximally-contrasting rhythmic temporal patterning in 1.5 Hz alternations \cite{petitto2001language,petitto2004baby}. Inside this temporal envelope were specific phonetic-syllabic contrasts, including 3 maximally-contrasting phonetic hand primes in ASL that human infants first begin to perceive and produce in language development: /5/, /B/, /G/ with contrastive transitions /B/=>/5/, /5/=>/F/, /G/=>/F/, plus allophonic variants. The Nursery Rhyme patterns were produced such that %(1) they represent maximally-contrasting phonetic units, (2) have maximally-contrasting rhythmic temporal frame and they also have (3) 
	each had baby-appropriate lexical meanings. % Each pattern consists of a series of fined grained meanings. 
	Below we provide an example of one of the four Nursery Rhymes, though each were designed with the same canonical structure.\\% (see also video clip)\\
	%This is a list of 4 ASL nursery rhymes with their respected action series: % Currently the avatar can produce 4 different nursery rhymes:% {(\color{red} Explain each of them!)}
	\textbf{BOAT (Phonetic-Syllabic units /B/, /5/ ) }
	\begin{enumerate}
		\item BOAT (/B/, palm in)
		\item BOAT-on-WATER (/B/+ movement modulation, palm up)
		\item WAVE (/5/+same movement modulation, palm down)
	\end{enumerate}
	\begin{comment}
	\begin{itemize}
		\item \textbf{BOAT (Phonetic-Syllabic units /B/, /5/ ) }
		\begin{enumerate}
			\item BOAT (/B/, palm in)
			\item BOAT-on-WATER (/B/+modulation, palm up)
			\item WAVE (/5/+SAME modulation, palm down)
		\end{enumerate}
		%	\textit{Sequence: (1) BOAT (/B/, palm in) (2) BOAT-on-WATER (/B/+modulation, palm up) (3) WAVE (/5/+SAME modulation, palm down). }
		
		\item \textbf{PIG (Phonetic-Syllabic unit: /5/)}
		\begin{enumerate}
			\item PIG (/5/, Chin)
			\item PET (/5/, called ``center space'' in Linguistic sign notation)
			\item HAPPY (/5/ + double-handed, Chest)
		\end{enumerate}
		\item \textbf{FISH (Phonetic-Syllabic unit: /B/ (allophonic))}
		\begin{enumerate}
			\item FISH (/B/, ``center space'')
			\item FINS (/B/+double-handed, Head)
			\item SWIMS (away) (/B/, Cross-Body)
		\end{enumerate}
		\item \textbf{CAT (Phonetic-Syllabic units: /5/; /G/allophonic; /BENT5/; and /F/} 
		\begin{enumerate}
			\item  Grandma has red cat  [/5/=>/G/] and [/G/=>/F/];
			\item  Grandma has white cat [/5/=>/BENT5/] and [/BENT5/=>/F/]
		\end{enumerate}
		
	\end{itemize}
	\end{comment}
\end{enumerate}

\begin{comment}
def triadic(self):
self.alyssa_lookAtMaki(waitTillFinish=False)
self.maki_LookatAlissa()
self.alyssa_nod(waitTilFinish=False)
self.maki_Nod()
self.alyssa_lookAtBaby(waitTillFinish=False)
self.maki_lookAtBaby()

\end{comment}

\subsubsection{\textbf{Action Sequences}}
We define an action sequence as a plan for a timed sequence of primitive actions by the agents that will be executed in order as planned. An example of an action sequence is the triad social greetings between the agents: (1) Avatar turns toward the Robot. (2) Robot turns toward the Avatar. (3) Avatar and Robot both nod to each other. (4) Avatar signs LOOK-AT-ME to both baby and Robot. (5) Avatar signs READY? to both baby and Robot. (6) Avatar turns back and looks at baby.
\begin{comment}
\begin{enumerate}
	\item Avatar turns toward the Robot
	\item Robot turns toward the Avatar
	\item Avatar and Robot both nod to each other
	\item Avatar signs LOOK-AT-ME to both baby and Robot 
	\item Avatar signs READY? to both baby and Robot 
	\item Avatar turns back and looks at baby
\end{enumerate}
\end{comment} 
%One use of this triad is when the baby is not looking and the agents will try to gain his/her attention by starting an interaction 

Another example is the familiarization sequence which is executed at the beginning of the experiment and will be described in detail in section \ref{sec:protocol}.  
%As explained in the next section, the dialogue controller will occasionally interrupt an action sequence in order to react in a more timely fashion to observed baby behaviors.

\begin{figure*}
	
	\centering
	\resizebox{16.1cm}{!}{

		\begin{forest}
			%	sn edges,
			%	l sep+=3em
			[\large{\textbf{State}} 
			[\normalsize{\textbf{AOI:In-Between}}, l=1.5cm [\normalsize{\textbf{Parasympathetic}} [\small{\framebox{\Longstack[l]{Avatar tries to gain\\ attention, then nursery\\ rhyme episode begins}}}, tier=first] ]
			[\normalsize{\textbf{Sympathetic}} [\small{\framebox{\Longstack[l]{Question solicitation from \\Avatar Robot is engaging \\with her}}},tier=first ]]]
			[\normalsize{\textbf{AOI:Robot/Avatar}}, tier=first, l=4cm [\normalsize{\textbf{BB:Vegetative/Crying/Fussing}}, l=1.5cm [\normalsize{\textbf{AOI:Robot}}, l=1.2cm [\small{\framebox{\Longstack[l]{Robot engages with baby \\ tries to shift baby's gaze\\ toward Avatar }}}, tier=second ]] 
			[\normalsize{\textbf{AOI:Avatar}},l=1.2cm [\small{\framebox{\Longstack[l]{Avatar engages with baby\\ What?/What's wrong/Peekaboo }}}, tier=second ]] ]
			[\fcolorbox{white}{white}{\parbox{ 0.27\linewidth-1\fboxsep-2\fboxrule}{%
					\centering \normalsize{\textbf{BB:Attention/Signs/\\Waving/Babbling/\\Reaching/Pointing}}}}, l=1.8cm           				
			[\normalsize{\textbf{AOI:Robot}}  [\small{\framebox{\Longstack[l]{Robot engages with baby \\ turns to Avatar, Avatar \\tries to gain attention }}}, tier=last ]] 
			[\normalsize{\textbf{AOI:Avatar}} [\normalsize{\textbf{Parasympathetic}}, tier=second [\small{\framebox{\Longstack[l]{Nursery Rhyme\\ episode begins }}}, tier=last ]]
			[\normalsize{\textbf{Sympathetic}},tier=second [\small{\framebox{\Longstack[l]{Social Routines and\\ question solicitation\\ from Avatar }}}, tier=last ]]]]
			]
			[\normalsize{\textbf{AOI:Outside}}, l=1.5cm, s=5cm [\normalsize{\textbf{Parasympathetic}} [\small{\framebox{\Longstack[l]{Avatar tries to gain\\ attention, if successful,\\ nursery rhyme episode\\ begins}}},tier=first] ]
			[\normalsize{\textbf{Sympathetic}} [\small{\framebox{\Longstack[l]{Avatar responding, Robot\\ looking at her and\\ occasional nodding}}},tier=first ]]]]                                             
		\end{forest}
	}
	\caption{Summarized decision tree based on system variables}
	\label{fig:decisionTree}
	\vspace{-3mm}
\end{figure*}
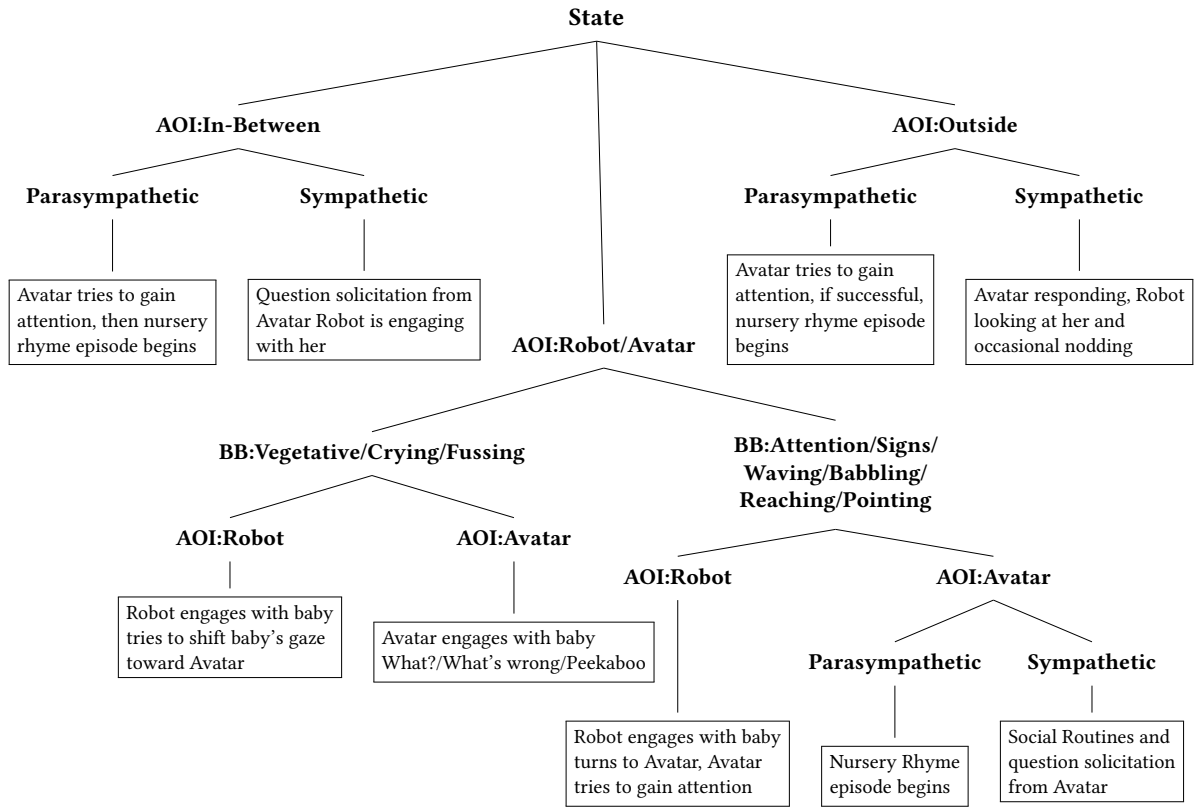

\begin{comment}
\subsubsection{\textbf{Dialog Interactions}}
Dialog policy is a higher level control over the action sequences
based on the present state of the interaction. It takes into account
the precedence of the signals and determines the action sequences. 
After each primitive action, the system checks for updates in state variables. Of all of the input signals, baby-behavior signals can interrupt the current plan. As if we observe any changes in baby's social behavior, we update our plan accordingly to maintain a socially contingent interaction. Also if there is a malfunction in either of the agents, they will recover their state by going back to the their idle behavior. More discussion about the policy will be provided in section \ref{sec:protocol}.

\begin{enumerate}
\item \textbf{Primitive Behaviors}: Defined as atomic actions of the agents which are single behaviors such as nodding or a single nursery rhyme.
\item \textbf{Action Sequences}: A trail of actions by the agents that will be executed in order as planned 
\item \textbf{Dialog Interactions}: Higher level control over the action sequences based on the present state of the interaction
\end{enumerate}

These behaviors are triggered in synchronization and coordination with each other to conduct a coherent and contingent interaction. We will explain more about the policy used to determine the output signals in the section \ref{sec:protocol}. 

\end{comment}

\subsection{Interaction Protocol}
\label{sec:protocol}
%what information state variables are tracked - how they are updated by perception, and how these are used to decide what the virtual human and robot should do.
%During human interactions, all parties have their own plan from each point forward based on the current state and the context of the interaction. Each party acts according to his plan to reach his own goals in the conversation, but plans constantly change during the course of the interactions to adapt to the other parties' actions and other surrounding variables. Similarly, for our 3 way interaction, 
At each point in the 3 way robot-avatar-baby interaction, the system has a sequence of actions as the current plan for the agents to execute. These 
%are  relatively long plans which 
are designed with the assumption that the baby will behave accordingly, but
% to them. 
%But in the course of the execution,
 if the baby acts differently, the planned actions may get removed or get updated by a completely different plan, in order 
%depending on the baby's behaviors and his role in the interaction. 
%So the system is continuously looking for updates in baby's behaviors and tries to adapt its plan in real time 
to maintain a socially contingent interaction. The only signal that causes an interruption in the execution of currently planned actions is the input signal coming from the baby behavior interface. In this case, the policy overrides the current plan with a new plan according to the new state of the baby. 

Each set of input combinations leads to a sequence of actions from the Avatar and Robot. %which will be executed in order
 %unless interrupted by baby behavior leading to %overwritten bya new set of actions. 
Theoretically speaking, considering only the 3 input variables coming from the perceptual components, we are looking at an information state space of $4 * 5 * 23 = 460$ possible combinations. 
However, not every combination is possible or likely to happen; but to build a completely reliable system all combinations should be considered. 
We used a rule based policy which will trigger specific sequences of behaviors based on predefined combination of variables. Figure \ref{fig:decisionTree} shows a highly abstract decision tree used as part of the policy in which many branches are aggregated with each other. %Also, e
Each branch consists of more fine-grained branches based on different input values for the baby-behavior, fixation of gaze, former executed plans and other state variables. 
%Considering the space constraints, depicting the whole decision tree was not feasible in this paper but the illustrated tree shows the high level strategy plan of the system. 

In order to make the baby familiar and comfortable with its surrounding, we begin the experiment with a familiarization episode. The goal of this period is to introduce the agents as conversational partners and make the baby feel involved in this multiparty interaction. This is a trace of what happens in the familiarization episode: 
\begin{comment}
\begin{enumerate}
	\item Robot wakes up from his sleeping position \label{wakeup}
	\item Robot nods towards the baby \label{nodtobaby}
	\item Robot turns to Avatar \label{robot_turn_to_avatar} 
	\item Avatar looks at the Robot
	\item Avatar nods to the Robot \label{avatar_nods_to_robot}
	\item Avatar turns toward the Baby \label{avatar_turn_to_baby}
	\item Avatar waves at the Baby \label{wave}
	\item Avatar signs good morning \label{goodmorning}
	\item Robot turns back to the Baby 
\end{enumerate}
\end{comment}
At the start of the experiment, both Avatar and the Robot are in their idle and neutral form. The Robot's head is down with its eyes closed and Avatar is standing still looking forward. Robot wakes up from his sleeping position, sees the baby and nods as an indication that it has acknowledged the baby's presence and then turns to Avatar. Avatar looks at the robot and acknowledge it by nodding and then turns to the baby. Then the Avatar tries to gain the baby's attention by waving to it. The Avatar will sign GOOD MORNING toward both the Baby and the Robot to begin the interaction.

%You can see a sample trace of the system actions in
 Table \ref{tab:sequence} shows a sequence of snapshots drawn from one interaction with a baby, along with different state variable values and an explanation of the state of the system at each point. This triad interaction between Avatar-Robot-Baby consists of the agents greeting each other (as participants in the conversation) and then the Avatar taking the floor and signing to the baby. Robot will nod to the Avatar occasionally to establish his role as a 3rd party conversationalist. We call this sequence a ``Nursery Rhyme episode''.
%Consider the following example: \\
%Baby is in a babbling state, looking at robot and the thermal signal indicate that the baby is in a parasympathetic state. Assume that both Robot and Avatar are in their idle form which is facing the baby. In this case, we will begin a nursery rhyme episode; Robot will look at the avatar, Avatar will turn to the robot to acknowledge him as a partner in conversation and nods to him, then the avatar will turn back to the baby and will begin to provide language stimuli and signing nursery rhymes to the baby. And the end of the nursery rhyme they both go back to the idle state as in they would do at the end of an utterance in an interaction. Figure ? shows a sequence of snapshots of the system going through this sequence of behaviors with the assumption that this sequence is not interrupted through the end.  ({\color{red} insert snapshots of the dialogue trace! a lot of space :DDD })
%({ \color{blue} Talk about the decision tree of the protocol , precedence of the signals, / sketch the tree }) 
%the system checks for updates to the variable values every 10 ms to maintain a responsive and contingent conversation. 

\begin{table*}[b]
	
	\centering
	\resizebox{18cm}{!}{
	%{\renewcommand\arraystretch{2.4} \setlength\minrowclearance{1.2pt}
	\begin{tabular}{|m{0.1in}|c|m{0.35in}|m{1.2in}|m{2.4in}|m{1.9in}|}
		%\begin{tabular}{|m{0.35in}|m{1.6in}|m{10cm}| }
		
		\rowcolor[HTML]{C0C0C0} \hline
		No. & AOI & Thermal &  Baby & System & Description \\ \hline
		1 & Between & + & 
		\vspace{1mm} \includegraphics[width=1.2in,height=1.2in]{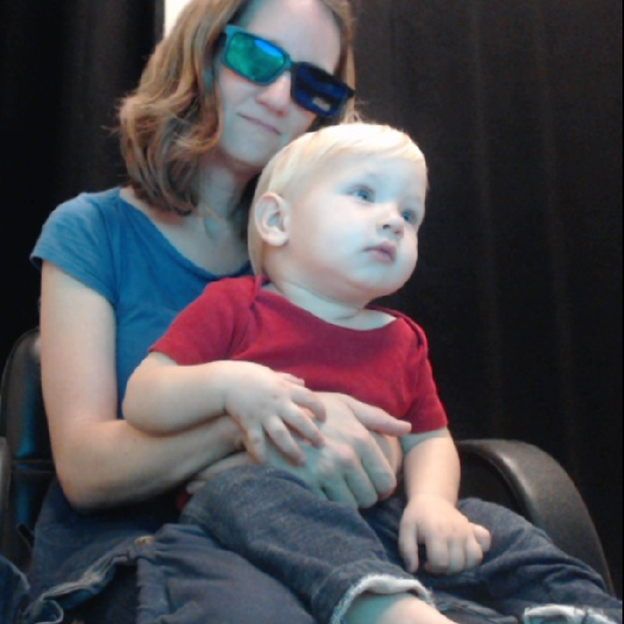} & \vspace{1mm}\includegraphics[width=2.4in]{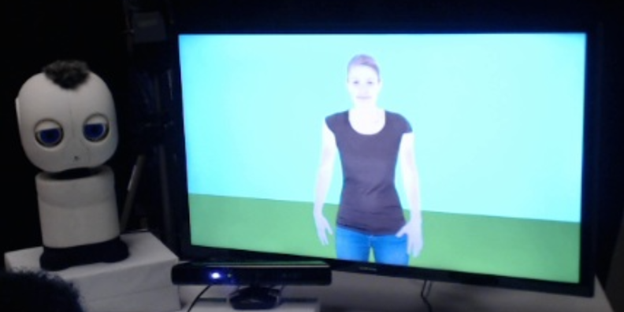} & Baby is focusing, paying attention to the system. He is looking somewhere in between avatar/ robot. The goal of the system is to shift his gaze toward the Avatar\\ \hline 
		2 & Avatar & ++ & \vspace{1mm} \includegraphics[width=1.2in,height=1.2in]{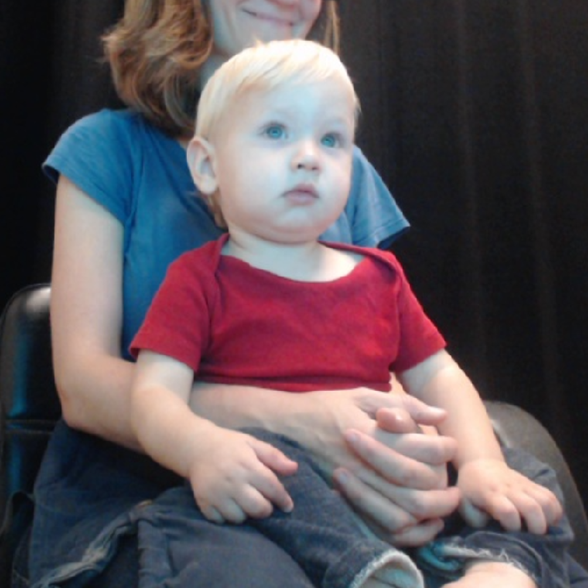}  & \vspace{1mm} \includegraphics[width=2.4in]{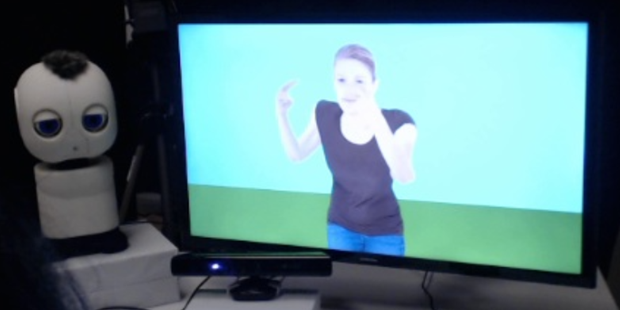} & Avatar tries to gain attention by signing LOOK-AT-ME. Thermal signal indicates that baby is in an engaged prosocial (parasympathetic) state (or ``ready to learn''), thus the system transits to a nursery rhyme episode. \\ \hline
		%3 & Avatar & ++ & \vspace{1mm} \includegraphics[width=1.2in,height=1.2in]{Figures/baby_alissa2}  & \vspace{1mm} \includegraphics[width=2.4in]{Figures/looking_at_eachOther} & Avatar and robot turn to each other. \\ \hline 
		3 & Robot & ++ & \vspace{1mm} \includegraphics[width=1.2in,height=1.2in]{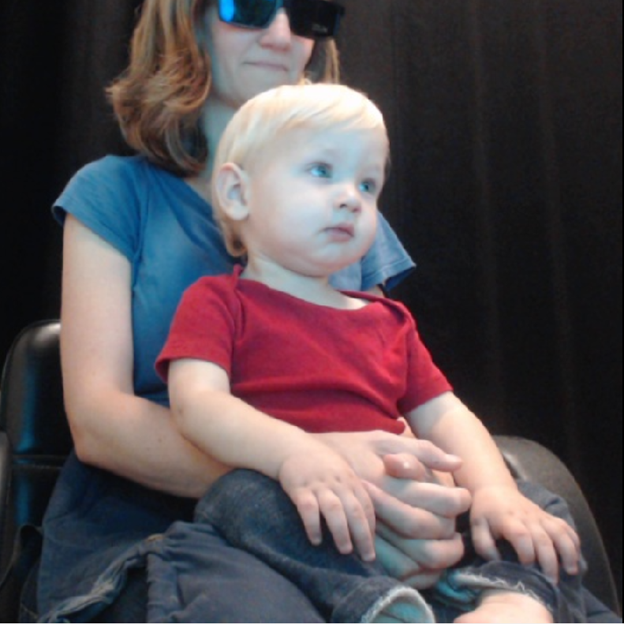} & \vspace{1mm} \includegraphics[width=2.4in]{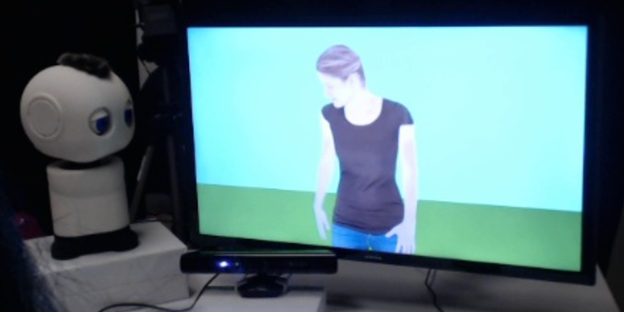} & The agents nod after turning to each other. The goal is for the Avatar to acknowledge the robot as a 3rd party conversationalist in the interaction before she takes the floor and begin signing. \\ \hline 
		4 & Avatar & ++ & \vspace{1mm} \includegraphics[width=1.2in,height=1.2in]{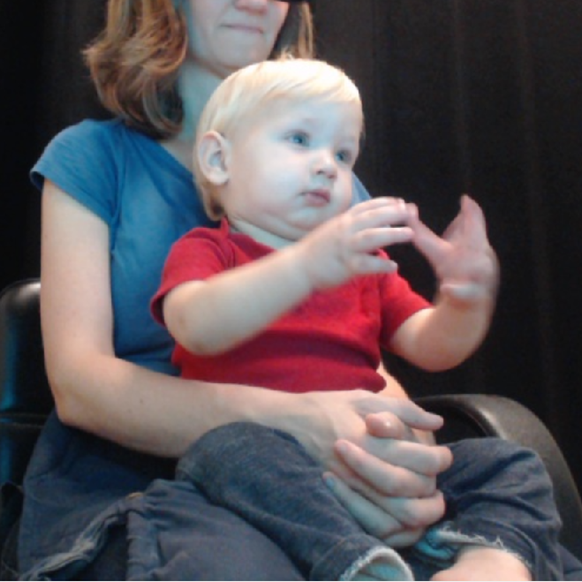} &\vspace{1mm} \includegraphics[width=2.4in]{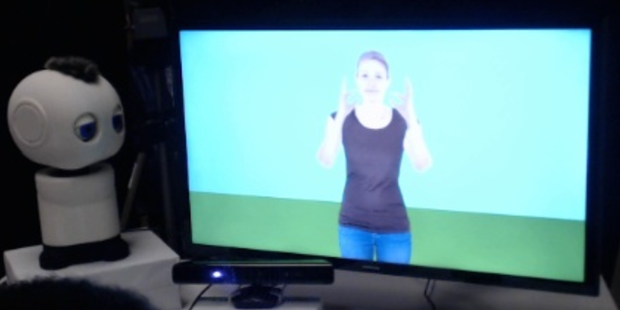} & Avatar begins a nursery Rhyme. Robot turns to her in the middle and nods towards her. Baby is copying the Avatar and is producing signs/proto-signs in response to the Avatar's linguistic input.  \\ \hline 
		5 & Neither & None & \vspace{1mm} \includegraphics[width=1.2in,height=1.2in]{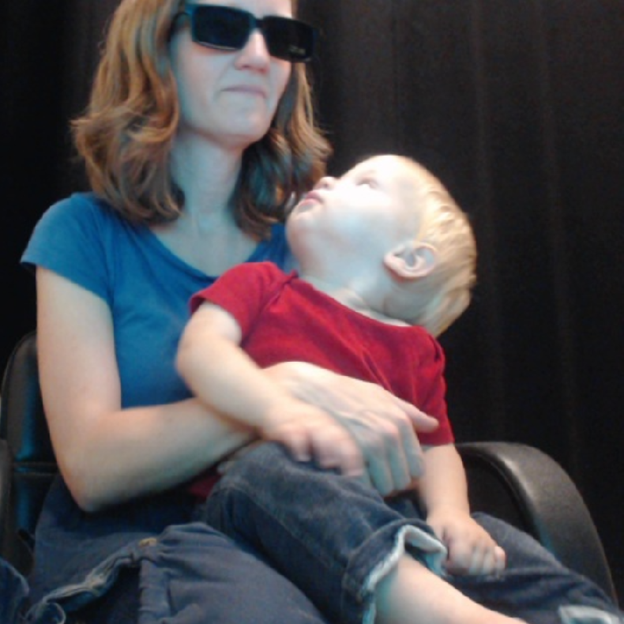} &\vspace{1mm} \includegraphics[width=2.4in]{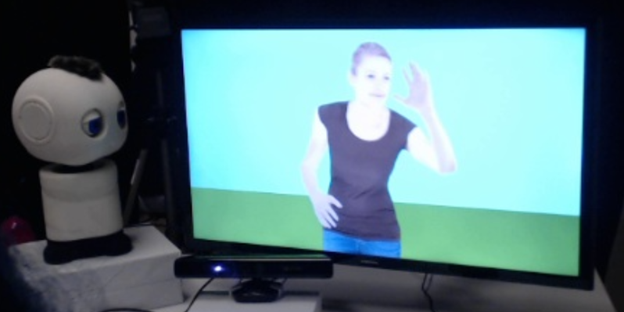} & Baby turns to look at his mom to exhibit the classic social referencing behavior. Avatar signs ATTEND-TO-ME at the baby and tries to get back his attention. \\ \hline

	\end{tabular}
	}
	\caption{A sequence of snapshots drawn from a sample experiment showing different stages of the interaction. (Participant is a hearing male with no sign exposure and is 12 months and 1 day old)}
	\label{tab:sequence}

\end{table*}

	\section{Evaluation}
	\label{sec:evaluation}
	The final dialogue manager routines described in the previous section were tested in interaction with 8 babies: 2 females and 6 males with average age of 9 months and 20 days (range 7-13 months) from whom 5 were hearing participants not sign-exposed, two hearing sign-exposed and one deaf sign-exposed. %The experimental protocol included a calibration and introduction routine with an experimenter.
On first entering the room, babies were permitted to briefly touch the robot in a quick greeting interaction with the agents, as had been shown to be important in older children ($\sim$18 months old) interacting with robots \cite{meltzoff2010social}. Following this brief greeting exchange lasting less than 1 minute, the babies were seated on their parent's lap, facing the Avatar and Robot, as shown in Figure~\ref{fig:overview}. 
Parents were instructed to wear sunglasses throughout the experiment to block the eye-tracker from registering their eyes.
%As an experimental design feature, parents wore sunglasses throughout the experiment
%% which was an intentional design feature meant 
%to block the eye-tracker from registering their eyes.
%%recording eye-tracking artifacts from 
%%the parent's eyes. 
The session was divided into two conditions: (1) The parent did not participant in the agent-baby interaction (2-way); (2) the parent joined the agent-baby interaction (3-way). This provides additional opportunity for social referencing and conversational scaffolding involving Avatar-Baby-Parent triads. 

%Whether young babies would be positively engaged with the system constituted an important scientific question. 
We first asked the important scientific question as to whether RAVE can engage the infants' attention. Perhaps these young babies would not see the agents as interesting social interlocutors as predicted, but possibly boring objects, or worse, a source of agitation. When confronted with unknown humans and/or situations of novelty, babies at this age are prone to {\em stranger anxiety}~\cite{greenberg1973infant}. Given that our 8 babies (age range 7-13 months) were within the onset period of ``stranger anxiety'' (onset range 6-12 months), crying and fussing could have occurred. Thus, upon the babies'  first contact with the Robot-Avatar system, the babies could have been interested in the agents, but they also could have fussed, become distracted, etc., at which point the interactional session would have been immediately ceased. None of these distracted behaviors were observed and instead, all 8 of the tested babies exhibited positive engagement behaviors, including:
\begin{comment}
Whether young babies would be positively engaged with the system constituted an important scientific question. Perhaps the baby would not see the agents as interesting social interlocutors as predicted, but possibly boring objects, or worse, a source of agitation. When confronted with unknown humans and/or situations of novelty, babies at this age are prone to {\em stranger anxiety}~\cite{greenberg1973infant}. %Given that our 8 babies were within the peak period of ``stranger anxiety'', crying and fussing were a concern, as the peak stranger anxiety period occurs within the same ages as our babies: approximate range 6-12 months.
Given that our 8 babies were within the period of ``Stranger Anxiety'' (approximate range 6-12 months), crying and fussing could have occurred. Thus, upon the babies' first contact with the robot-avatar system, a variety of behavioral responses were indeed possible. The babies could have been interested in the agents, ut they also could have fussed, become distracted, etc., at which point the interactional session would have been instantly ceased. None of these distracted behaviors were observed and instead, all 8 of the tested babies exhibited positive engagement behaviors, including: 
\end{comment}
(1) Immediate visual engagement (locked attention) with the agents, (2) Sustained attention (persisting over time) and (3) Visually tracked (gaze following) the Avatar and Robot as they interacted with each other and the baby.
\begin{comment}
\looseness -1
\vspace{-2mm}
\begin{enumerate}
\item Immediate visual engagement (locked attention) with the agents.
\item Sustained attention (persisting over time). %typically $\sim$4 minutes, with some babies sustained in attention as long as 6 minutes -  this is an atypical %attention engagement window for very young infants as were these babies.) 
\item Visually tracked (gaze following) the avatar and robot as they  interacted with each other and the baby.
\end{enumerate} 
\end{comment}

%Most babies (6 out of 8) exhibited sustained attention lasting $\sim$5 minutes (average 5m26s, range 4m12s-7m47s). % with some sustained in attention as long as 7 minutes. 
%This is an atypical attention engagement window for very young infants. % The remaining sessions involving 2 of the 8 babies were ended by us at approximately 2-3 minutes into the session due to fussiness, and one baby cried. But these 2 babies came in fussy, changed to riveted attention upon sight of the agents, and then slipped into a fussy state again. The fact that the presence of the agents halted the babies' preexisting emotional agitation for several minutes is in itself remarkable, and invites us to understand why was it so.
%Even the 2 babies who came into the experiment fussy were able to participate for  $\sim$2 minutes (average 2m49s, range 2m13s-3m25s) and changed to riveted attention upon sight of the agents, and then slipped into a fussy state again thus we decided to terminate the session. 
All 8 babies exhibited sustained engagement with RAVE lasting 4-5 minutes (average 3m40s; range 1m33s-4m56s).
This is an atypical attention engagement window for very young infants. There was only one baby with the low engagement time (1m33s; > 2 standard deviations from the mean), but this baby entered the room very fussy. Although fussy on entering, she changed to riveted attention upon sight of the agents, and then slipped into a fussy state again at which point, we terminated the session. If we were to remove this outlier, the average sustained engagement time for the remaining babies is nearly 4 minutes (3m58s). Interestingly, this baby was an outlier for another reason. Her age (13mths;16days) fell outside of our predicted window of peak infant engagement for RAVE (age range 6-12 months). Her performance thus provided preliminary support for our prediction that RAVE was most optimal for babies within the developmental period when they had peaked sensitivity to the rhythmic patterning of language, ages 6-12 months. The fact that the presence of the agents impacted all babies' preexisting emotional and/or attentional states for such durations is in itself remarkable, and invites us to understand why was it so.  
%Even the 2 babies who came into the experiment fussy were able to engage in sustained attention and participate for over 2 minutes (2m13s, 3m25s).
%They changed to riveted attention upon sight of the agents, and then slipped into a fussy state again at which point, we terminated the session. The fact that the presence of the agents halted the babies' preexisting emotional agitation for several minutes is in itself remarkable, and invites us to understand why was it so.

%These engagement behaviors occurred with all babies,
%We observed such sustained visual attention in all babies, even hearing babies with no prior exposure to signed language, with interesting group differences.% We found that early Exposure to Sign Language has greater impact on visual attention span as opposed to babies who had no Sign Exposure (Non Sign Exposed), and, Early dual language exposure (ASL-English) has greatest combined positive impact on visual attention span (=longest visual engagement with RAVE).
We observed such sustained engagement in all babies, even hearing babies with no prior exposure to signed language, meaning that something about the avatar's productions was engaging to the babies even though they could not understand the meanings of the signs, with interesting group differences. For example, we found that our one baby with early bilingual language exposure (i.e., early ASL and early English exposure) had the greatest combined positive impact on its engagement span (longest experiment run time of 4m56s). This finding corroborates our earlier studies showing significant processing advantages afforded to babies and children with early bilingual language exposure \cite{petitto2012perceptual}. 
%We found that early exposure to signed language had greater impact on visual attention span as opposed to babies who had no exposure to a signed language (Non Sign Exposed), and early dual language exposure (bilingual ASL-English exposure) in one baby had the greatest combined positive impact on visual attention span (=longest visual engagement with RAVE). 

% We also observed two broad classes of baby reactions. One set of babies was visually transfixed, so much so they watched riveted with relatively few social interaction solicitations back to the agents. Another set of babies was more active, showing not just engagement and visual gaze tracking of the agent behaviors, but also engaging in contingent social interactions with the agents. 
%Indeed, some babies produced robust social interaction solicitations to the agents (e.g., waving HI, pointing, reaching, etc.), and remarkably attempts to copy the agents, especially the Avatar's sign productions. 
The second condition, where parents were permitted to join in as they would naturally, allows for baby's social referencing to be acknowledged and responded to. In fact, we also observed instances where (nonsigning) parents copied the Avatar's signs and encouraged the baby to react and interact with the Avatar (only that will be picked up by the Avatar to continue its cycles).
 %We observed rich social referencing from the baby to the parent and back to the agents. 

%We also performed a detailed case-study with one of the babies (7 months old hearing sign-exposed male). 
Our second question was whether the artificial agents can elicit socially interactive and socially contingent conversation with an infant, above and beyond babies' production of prosocial emotional engagement/sustained visual attention. In an intensive analysis of 4 of the 8 babies (as analyses involve hundreds of hours of frame-by-frame video transcription with trained experts, behavioral coding, and reliability checks), all four babies produced social interactions and/or solicitations to the agents (e.g., waving HI, pointing, reaching, etc.) and attempted to copy the avatar, either through attempts to copy the avatar's signs (and components of signs) or matching the avatar's rhythmic movements at the nucleus of its sign productions. This novel finding is noteworthy because most babies (3 of 4) were never exposed to signed language yet attempted to copy the Avatar's linguistic signed language productions, and as noted above, they did so without understanding the meaning of the avatar's signs.  
%Most all babies produced social interactions and/or solicitations to the agents (e.g., waving HI, pointing, reaching, etc.). Remarkably, some babies never exposed to signed language attempted to copy the Avatar's linguistic signed language productions. Of note, the babies produced the largest percentage of spontaneous behavioral responses to the avatar (61\%) over the robot (54\%) and demonstrated implicit/tacit knowledge of the avatar's differential communicative states (e.g., the babies' behavioral response rates differed depending on whether the avatar was in active linguistic in contrast to when in non-active idle productions). 
Crucially, the babies' powerful engagement with the avatar occurred even though the avatar is an % inanimate
 artificial agent on a flat TV monitor. %To gain the greatest insight into this question, 
We also performed a detailed case-study with one of the babies (a 7 month-old hearing baby boy who was exposed to signed language/ASL). In particular, we examined: (1) Whether the baby performed age-appropriate proto-linguistic behavior? (2) Whether this was produced in a socially contingent sequence as solicited by the Avatar's linguistic behavior?
%\begin{enumerate}
%	\item Whether the baby performed age-appropriate proto-linguistic behavior?
%	\item  Whether this was produced in a socially contingent sequence as solicited by the Avatar's linguistic behavior?
%\end{enumerate}

In pursuit of these questions, we first coded the videos of conversational interactions with respect to Avatar and baby behaviors, followed by reliability checks. The rigorous coding was done by trained coders in the field of child language. Every video was coded for the categories of social conversational turns and content (see Figure \ref{fig:babyWoz}) along with the time marking in coding and total time length of coded segments. 
Regarding question (1), we see linguistic behavior from the baby in both conditions (with and without the parent joining in). The baby waved and produced proto-signs related two distinct Nursery Rhymes. Regarding question (2), the sign-productions in all cases appeared as socially contingent reactions to the Avatar. Baby proto-signs were produced within a few seconds of the Avatar producing the relevant Nursery Rhymes. Baby social behaviors, such as waving, were produced as a response to social routines such as the signs for HELLO or GOODBYE. Thus, we see that the agents performing dialogue routines, in reaction to continuous multimodal sensory updates, were successful in soliciting socially contingent conversation from the infant. 
This would suggest the potential viability for using this kind of system for language learning in young infants. 

		\balance
	\section{Conclusion and Future Work }
	\label{sec:conclusion}
	While our system shows much potential, there is still much to be done to achieve the ultimate goal of facilitating an infant to learn visual sign language. We have established that the system can engage infants and stimulate socially contingent rudimentary conversation using the rhythmic-temporal patterned language stimuli and the dyadic and multiparty social interactions. One strand of future work involves making the system more autonomous and robust. We would like to replace the observer GUI with automated perception of conversationally relevant baby behaviors, by training  behavior recognition models using collected Kinect and video data.

We also want to streamline the hardware footprint and start to look at whether the system can be used repeatedly, outside the laboratory, such that persistent learning over time can be achieved and assessed. Also, the system can adapt itself to behave differently to each baby to accommodate to specific behavioral and learning patterns of each individual. Another focus is to extend the dialogue routines to focus more specifically on the critical Agent-Infant-Parent triad. In particular, we desire to look at whether even parents who don't previously know sign language can assist in the child's learning (and learn themselves). This work has the potential for vast societal benefits, given the potential of baby's interaction with this type of system to ``wedge open'' their language learning capacity during the critical period of phonetic-syllabic development until the baby can receive systematic language input \cite{petitto2012perceptual}. 
	\begin{acks}
		The primary funding for this research was from the W.M. Keck Foundation (Petitto, PI: ``Seeing the Rhythmic Temporal Beats of Human Language''), the National Science Foundation ``INSPIRE'' (Petitto, PI: IIS-1547178, ``The RAVE Revolution for Children with Minimal Language Experience During Sensitive Periods of Brain and Language Development''), and the National Science Foundation Science of Learning Center Grant at Gallaudet University (Petitto, Co-PI and Science Director: SBE 1041725, ``Visual Language and Visual Learning, VL2''). We extend sincere thanks to our colleagues on the project, including Ari Shapiro, Andrew Feng, Brian  Scassellati, Jake Brawer, Katherine Tsui, Melissa Malzkuhn, Jason Lamberton, Adam Stone, Geo Kartheiser, and Gallaudet University student research assistants Rachel Sortino, Kailyn Aaron-Lozano, and Crystal Padilla. We especially thank the families who so generously gave of their time and support to participate in this study. 	
	\end{acks}
	
	\balance
	\clearpage
	\balance
	\bibliographystyle{ACM-Reference-Format}
	\bibliography{RAVE-bibliography}


\begin{thebibliography}{49}

%%% the Web address.

\ifx \showCODEN    \undefined \def \showCODEN     #1{\unskip}     \fi
\ifx \showDOI      \undefined \def \showDOI       #1{#1}\fi
\ifx \showISBNx    \undefined \def \showISBNx     #1{\unskip}     \fi
\ifx \showISBNxiii \undefined \def \showISBNxiii  #1{\unskip}     \fi
\ifx \showISSN     \undefined \def \showISSN      #1{\unskip}     \fi
\ifx \showLCCN     \undefined \def \showLCCN      #1{\unskip}     \fi
\ifx \shownote     \undefined \def \shownote      #1{#1}          \fi
\ifx \showarticletitle \undefined \def \showarticletitle #1{#1}   \fi
\ifx \showURL      \undefined \def \showURL       {\relax}        \fi
% invisible to TeX
\providecommand\bibfield[2]{#2}
\providecommand\bibinfo[2]{#2}
\providecommand\natexlab[1]{#1}
\providecommand\showeprint[2][]{arXiv:#2}

\bibitem[\protect\citeauthoryear{ActiveMQ}{ActiveMQ}{2018}]%
        {activeMQ}
ActiveMQ \bibinfo{year}{2018}\natexlab{}.
\newblock \bibinfo{title}{Apache ACTIVEMQ}.
\newblock \bibinfo{howpublished}{\url{http://activemq.apache.org}}.
\newblock
\newblock
\shownote{Accessed: 2018-04-25.}


\bibitem[\protect\citeauthoryear{Aran, Ari, Akarun, Sankur, Benoit, Caplier,
  Campr, Carrillo, and Fanard}{Aran et~al\mbox{.}}{2009}]%
        {aran2009signtutor}
\bibfield{author}{\bibinfo{person}{Oya Aran}, \bibinfo{person}{Ismail Ari},
  \bibinfo{person}{Lale Akarun}, \bibinfo{person}{B{\"u}lent Sankur},
  \bibinfo{person}{Alexandre Benoit}, \bibinfo{person}{Alice Caplier},
  \bibinfo{person}{Pavel Campr}, \bibinfo{person}{Ana~Huerta Carrillo}, {and}
  \bibinfo{person}{Fran{\c{c}}ois-Xavier Fanard}.}
  \bibinfo{year}{2009}\natexlab{}.
\newblock \showarticletitle{SignTutor: An Interactive System for Sign Language
  Tutoring.}
\newblock \bibinfo{journal}{\emph{IEEE MultiMedia}} \bibinfo{volume}{16},
  \bibinfo{number}{1} (\bibinfo{year}{2009}), \bibinfo{pages}{81--93}.
\newblock


\bibitem[\protect\citeauthoryear{Arita, Hiraki, Kanda, and Ishiguro}{Arita
  et~al\mbox{.}}{2005}]%
        {arita2005can}
\bibfield{author}{\bibinfo{person}{Akiko Arita}, \bibinfo{person}{Kazuo
  Hiraki}, \bibinfo{person}{Takayuki Kanda}, {and} \bibinfo{person}{Hiroshi
  Ishiguro}.} \bibinfo{year}{2005}\natexlab{}.
\newblock \showarticletitle{Can we talk to robots? Ten-month-old infants
  expected interactive humanoid robots to be talked to by persons}.
\newblock \bibinfo{journal}{\emph{Cognition}} \bibinfo{volume}{95},
  \bibinfo{number}{3} (\bibinfo{year}{2005}), \bibinfo{pages}{B49--B57}.
\newblock


\bibitem[\protect\citeauthoryear{Debevec}{Debevec}{2012}]%
        {debevec2012light}
\bibfield{author}{\bibinfo{person}{Paul Debevec}.}
  \bibinfo{year}{2012}\natexlab{}.
\newblock \showarticletitle{The light stages and their applications to
  photoreal digital actors}.
\newblock \bibinfo{journal}{\emph{SIGGRAPH Asia}} \bibinfo{volume}{2},
  \bibinfo{number}{4} (\bibinfo{year}{2012}).
\newblock


\bibitem[\protect\citeauthoryear{Finn}{Finn}{2010}]%
        {finn2010sensitive}
\bibfield{author}{\bibinfo{person}{Amy~Sue Finn}.}
  \bibinfo{year}{2010}\natexlab{}.
\newblock \bibinfo{booktitle}{\emph{The sensitive period for language
  acquisition: The role of age related differences in cognitive and neural
  function}}.
\newblock \bibinfo{publisher}{University of California, Berkeley}.
\newblock


\bibitem[\protect\citeauthoryear{Greenberg, Hillman, and Grice}{Greenberg
  et~al\mbox{.}}{1973}]%
        {greenberg1973infant}
\bibfield{author}{\bibinfo{person}{David~J Greenberg}, \bibinfo{person}{Donald
  Hillman}, {and} \bibinfo{person}{Dean Grice}.}
  \bibinfo{year}{1973}\natexlab{}.
\newblock \showarticletitle{Infant and stranger variables related to stranger
  anxiety in the first year of life.}
\newblock \bibinfo{journal}{\emph{Developmental Psychology}}
  \bibinfo{volume}{9}, \bibinfo{number}{2} (\bibinfo{year}{1973}),
  \bibinfo{pages}{207}.
\newblock


\bibitem[\protect\citeauthoryear{Higgins}{Higgins}{1980}]%
        {higgins1980outsiders}
\bibfield{author}{\bibinfo{person}{P Higgins}.}
  \bibinfo{year}{1980}\natexlab{}.
\newblock \showarticletitle{Outsiders in a hearing world}.
\newblock  (\bibinfo{year}{1980}).
\newblock


\bibitem[\protect\citeauthoryear{House}{House}{1976}]%
        {house1976cochlear}
\bibfield{author}{\bibinfo{person}{William~F House}.}
  \bibinfo{year}{1976}\natexlab{}.
\newblock \showarticletitle{Cochlear implants}.
\newblock \bibinfo{journal}{\emph{Annals of Otology, Rhinology \& Laryngology}}
  \bibinfo{volume}{85}, \bibinfo{number}{3} (\bibinfo{year}{1976}),
  \bibinfo{pages}{3--3}.
\newblock


\bibitem[\protect\citeauthoryear{Ioannou, Gallese, and Merla}{Ioannou
  et~al\mbox{.}}{2014}]%
        {ioannou2014thermal}
\bibfield{author}{\bibinfo{person}{Stephanos Ioannou},
  \bibinfo{person}{Vittorio Gallese}, {and} \bibinfo{person}{Arcangelo Merla}.}
  \bibinfo{year}{2014}\natexlab{}.
\newblock \showarticletitle{Thermal infrared imaging in psychophysiology:
  potentialities and limits}.
\newblock \bibinfo{journal}{\emph{Psychophysiology}} \bibinfo{volume}{51},
  \bibinfo{number}{10} (\bibinfo{year}{2014}), \bibinfo{pages}{951--963}.
\newblock


\bibitem[\protect\citeauthoryear{IR thermal camera}{IR thermal camera}{2018}]%
        {thermalImagingCamera}
IR thermal camera \bibinfo{year}{2018}\natexlab{}.
\newblock \bibinfo{title}{FLIR A655sc}.
\newblock \bibinfo{howpublished}{\url{https://www.flir.com/products/a655sc/}}.
\newblock
\newblock
\shownote{Accessed: 2018-04-25.}


\bibitem[\protect\citeauthoryear{Jaballah and Jemni}{Jaballah and
  Jemni}{2013}]%
        {jaballah2013review}
\bibfield{author}{\bibinfo{person}{Kabil Jaballah} {and}
  \bibinfo{person}{Mohamed Jemni}.} \bibinfo{year}{2013}\natexlab{}.
\newblock \showarticletitle{A Review on 3D signing avatars: Benefits, uses and
  challenges}.
\newblock \bibinfo{journal}{\emph{International Journal of Multimedia Data
  Engineering and Management (IJMDEM)}} \bibinfo{volume}{4},
  \bibinfo{number}{1} (\bibinfo{year}{2013}), \bibinfo{pages}{21--45}.
\newblock


\bibitem[\protect\citeauthoryear{Kipp, Heloir, and Nguyen}{Kipp
  et~al\mbox{.}}{2011}]%
        {kipp2011sign}
\bibfield{author}{\bibinfo{person}{Michael Kipp}, \bibinfo{person}{Alexis
  Heloir}, {and} \bibinfo{person}{Quan Nguyen}.}
  \bibinfo{year}{2011}\natexlab{}.
\newblock \showarticletitle{Sign language avatars: Animation and
  comprehensibility}. In \bibinfo{booktitle}{\emph{International Workshop on
  Intelligent Virtual Agents}}. Springer, \bibinfo{pages}{113--126}.
\newblock


\bibitem[\protect\citeauthoryear{Klima and Bellugi}{Klima and Bellugi}{1979}]%
        {edwards1979signs}
\bibfield{author}{\bibinfo{person}{Edward~S. Klima} {and}
  \bibinfo{person}{Ursula Bellugi}.} \bibinfo{year}{1979}\natexlab{}.
\newblock \bibinfo{title}{The signs of language}.
\newblock
\newblock


\bibitem[\protect\citeauthoryear{Kose, Akalin, and Uluer}{Kose
  et~al\mbox{.}}{2014}]%
        {kose2014socially}
\bibfield{author}{\bibinfo{person}{Hatice Kose}, \bibinfo{person}{Neziha
  Akalin}, {and} \bibinfo{person}{Pinar Uluer}.}
  \bibinfo{year}{2014}\natexlab{}.
\newblock \showarticletitle{Socially interactive robotic platforms as sign
  language tutors}.
\newblock \bibinfo{journal}{\emph{International Journal of Humanoid Robotics}}
  \bibinfo{volume}{11}, \bibinfo{number}{01} (\bibinfo{year}{2014}),
  \bibinfo{pages}{1450003}.
\newblock


\bibitem[\protect\citeauthoryear{Kose, Yorganci, Algan, and Syrdal}{Kose
  et~al\mbox{.}}{2012}]%
        {kose2012evaluation}
\bibfield{author}{\bibinfo{person}{Hatice Kose}, \bibinfo{person}{Rabia
  Yorganci}, \bibinfo{person}{Esra~H Algan}, {and} \bibinfo{person}{Dag~S
  Syrdal}.} \bibinfo{year}{2012}\natexlab{}.
\newblock \showarticletitle{Evaluation of the robot assisted sign language
  tutoring using video-based studies}.
\newblock \bibinfo{journal}{\emph{International Journal of Social Robotics}}
  \bibinfo{volume}{4}, \bibinfo{number}{3} (\bibinfo{year}{2012}),
  \bibinfo{pages}{273--283}.
\newblock


\bibitem[\protect\citeauthoryear{Kose, Yorganci, and Itauma}{Kose
  et~al\mbox{.}}{2011}]%
        {kose2011humanoid}
\bibfield{author}{\bibinfo{person}{Hatice Kose}, \bibinfo{person}{Rabia
  Yorganci}, {and} \bibinfo{person}{Itauma~I Itauma}.}
  \bibinfo{year}{2011}\natexlab{}.
\newblock \showarticletitle{Humanoid robot assisted interactive sign language
  tutoring game}. In \bibinfo{booktitle}{\emph{Robotics and Biomimetics
  (ROBIO), 2011 IEEE International Conference on}}. IEEE,
  \bibinfo{pages}{2247--2248}.
\newblock


\bibitem[\protect\citeauthoryear{Krcmar}{Krcmar}{2011}]%
        {krcmar2011word}
\bibfield{author}{\bibinfo{person}{Marina Krcmar}.}
  \bibinfo{year}{2011}\natexlab{}.
\newblock \showarticletitle{Word learning in very young children from
  infant-directed DVDs}.
\newblock \bibinfo{journal}{\emph{Journal of Communication}}
  \bibinfo{volume}{61}, \bibinfo{number}{4} (\bibinfo{year}{2011}),
  \bibinfo{pages}{780--794}.
\newblock


\bibitem[\protect\citeauthoryear{Krcmar, Grela, and Lin}{Krcmar
  et~al\mbox{.}}{2007}]%
        {krcmar2007can}
\bibfield{author}{\bibinfo{person}{Marina Krcmar}, \bibinfo{person}{Bernard
  Grela}, {and} \bibinfo{person}{Kirsten Lin}.}
  \bibinfo{year}{2007}\natexlab{}.
\newblock \showarticletitle{Can toddlers learn vocabulary from television? An
  experimental approach}.
\newblock \bibinfo{journal}{\emph{Media Psychology}} \bibinfo{volume}{10},
  \bibinfo{number}{1} (\bibinfo{year}{2007}), \bibinfo{pages}{41--63}.
\newblock


\bibitem[\protect\citeauthoryear{Kuhl}{Kuhl}{2004}]%
        {kuhl2004early}
\bibfield{author}{\bibinfo{person}{Patricia~K Kuhl}.}
  \bibinfo{year}{2004}\natexlab{}.
\newblock \showarticletitle{Early language acquisition: cracking the speech
  code}.
\newblock \bibinfo{journal}{\emph{Nature reviews neuroscience}}
  \bibinfo{volume}{5}, \bibinfo{number}{11} (\bibinfo{year}{2004}),
  \bibinfo{pages}{831}.
\newblock


\bibitem[\protect\citeauthoryear{Kuhl, Tsao, and Liu}{Kuhl
  et~al\mbox{.}}{2003}]%
        {Kuhl9096}
\bibfield{author}{\bibinfo{person}{Patricia~K. Kuhl},
  \bibinfo{person}{Feng-Ming Tsao}, {and} \bibinfo{person}{Huei-Mei Liu}.}
  \bibinfo{year}{2003}\natexlab{}.
\newblock \showarticletitle{Foreign-language experience in infancy: Effects of
  short-term exposure and social interaction on phonetic learning}.
\newblock \bibinfo{journal}{\emph{Proceedings of the National Academy of
  Sciences}} \bibinfo{volume}{100}, \bibinfo{number}{15}
  (\bibinfo{year}{2003}), \bibinfo{pages}{9096--9101}.
\newblock
\showISSN{0027-8424}
\showeprint{http://www.pnas.org/content/100/15/9096.full.pdf}


\bibitem[\protect\citeauthoryear{Leyzberg, Spaulding, Toneva, and
  Scassellati}{Leyzberg et~al\mbox{.}}{2012}]%
        {leyzberg2012physical}
\bibfield{author}{\bibinfo{person}{Daniel Leyzberg}, \bibinfo{person}{Samuel
  Spaulding}, \bibinfo{person}{Mariya Toneva}, {and} \bibinfo{person}{Brian
  Scassellati}.} \bibinfo{year}{2012}\natexlab{}.
\newblock \showarticletitle{The physical presence of a robot tutor increases
  cognitive learning gains}. In \bibinfo{booktitle}{\emph{Proceedings of the
  Annual Meeting of the Cognitive Science Society}}, Vol.~\bibinfo{volume}{34}.
\newblock


\bibitem[\protect\citeauthoryear{Manini, Cardone, Ebisch, Bafunno, Aureli, and
  Merla}{Manini et~al\mbox{.}}{2013}]%
        {manini2013mom}
\bibfield{author}{\bibinfo{person}{Barbara Manini}, \bibinfo{person}{Daniela
  Cardone}, \bibinfo{person}{Sjoerd Ebisch}, \bibinfo{person}{Daniela Bafunno},
  \bibinfo{person}{Tiziana Aureli}, {and} \bibinfo{person}{Arcangelo Merla}.}
  \bibinfo{year}{2013}\natexlab{}.
\newblock \showarticletitle{Mom feels what her child feels: thermal signatures
  of vicarious autonomic response while watching children in a stressful
  situation}.
\newblock \bibinfo{journal}{\emph{Frontiers in human neuroscience}}
  \bibinfo{volume}{7} (\bibinfo{year}{2013}), \bibinfo{pages}{299}.
\newblock


\bibitem[\protect\citeauthoryear{Meltzoff, Brooks, Shon, and Rao}{Meltzoff
  et~al\mbox{.}}{2010}]%
        {meltzoff2010social}
\bibfield{author}{\bibinfo{person}{Andrew~N Meltzoff}, \bibinfo{person}{Rechele
  Brooks}, \bibinfo{person}{Aaron~P Shon}, {and} \bibinfo{person}{Rajesh~PN
  Rao}.} \bibinfo{year}{2010}\natexlab{}.
\newblock \showarticletitle{"Social" robots are psychological agents for
  infants: A test of gaze following}.
\newblock \bibinfo{journal}{\emph{Neural networks}} \bibinfo{volume}{23},
  \bibinfo{number}{8-9} (\bibinfo{year}{2010}), \bibinfo{pages}{966--972}.
\newblock


\bibitem[\protect\citeauthoryear{Merla}{Merla}{2014}]%
        {merla2014thermal}
\bibfield{author}{\bibinfo{person}{Arcangelo Merla}.}
  \bibinfo{year}{2014}\natexlab{}.
\newblock \showarticletitle{Thermal expression of intersubjectivity offers new
  possibilities to human--machine and technologically mediated interactions}.
\newblock \bibinfo{journal}{\emph{Frontiers in psychology}}
  \bibinfo{volume}{5} (\bibinfo{year}{2014}), \bibinfo{pages}{802}.
\newblock


\bibitem[\protect\citeauthoryear{Payne}{Payne}{2018}]%
        {makiProject}
\bibfield{author}{\bibinfo{person}{Tim Payne}.}
  \bibinfo{year}{2018}\natexlab{}.
\newblock \bibinfo{title}{MAKI - A 3D Printable Humanoid Robot}.
\newblock
  \bibinfo{howpublished}{\url{https://www.kickstarter.com/projects/391398742/maki-a-3d-printable-humanoid-robot}}.
\newblock
\newblock
\shownote{Accessed: 2018-04-25.}


\bibitem[\protect\citeauthoryear{Petitto, Berens, Kovelman, Dubins, Jasinska,
  and Shalinsky}{Petitto et~al\mbox{.}}{2012}]%
        {petitto2012perceptual}
\bibfield{author}{\bibinfo{person}{Laura-Ann Petitto},
  \bibinfo{person}{Melody~S Berens}, \bibinfo{person}{Ioulia Kovelman},
  \bibinfo{person}{Matt~H Dubins}, \bibinfo{person}{K Jasinska}, {and}
  \bibinfo{person}{M Shalinsky}.} \bibinfo{year}{2012}\natexlab{}.
\newblock \showarticletitle{The ``Perceptual Wedge Hypothesis'' as the basis
  for bilingual babies' phonetic processing advantage: New insights from fNIRS
  brain imaging}.
\newblock \bibinfo{journal}{\emph{Brain and language}} \bibinfo{volume}{121},
  \bibinfo{number}{2} (\bibinfo{year}{2012}), \bibinfo{pages}{130--143}.
\newblock


\bibitem[\protect\citeauthoryear{Petitto, Holowka, Sergio, Levy, and
  Ostry}{Petitto et~al\mbox{.}}{2004}]%
        {petitto2004baby}
\bibfield{author}{\bibinfo{person}{Laura~Ann Petitto}, \bibinfo{person}{Siobhan
  Holowka}, \bibinfo{person}{Lauren~E Sergio}, \bibinfo{person}{Bronna Levy},
  {and} \bibinfo{person}{David~J Ostry}.} \bibinfo{year}{2004}\natexlab{}.
\newblock \showarticletitle{Baby hands that move to the rhythm of language:
  hearing babies acquiring sign languages babble silently on the hands}.
\newblock \bibinfo{journal}{\emph{Cognition}} \bibinfo{volume}{93},
  \bibinfo{number}{1} (\bibinfo{year}{2004}), \bibinfo{pages}{43--73}.
\newblock


\bibitem[\protect\citeauthoryear{Petitto, Holowka, Sergio, and Ostry}{Petitto
  et~al\mbox{.}}{2001}]%
        {petitto2001language}
\bibfield{author}{\bibinfo{person}{Laura~Ann Petitto}, \bibinfo{person}{Siobhan
  Holowka}, \bibinfo{person}{Lauren~E Sergio}, {and} \bibinfo{person}{David
  Ostry}.} \bibinfo{year}{2001}\natexlab{}.
\newblock \showarticletitle{Language rhythms in baby hand movements}.
\newblock \bibinfo{journal}{\emph{Nature}} \bibinfo{volume}{413},
  \bibinfo{number}{6851} (\bibinfo{year}{2001}), \bibinfo{pages}{35}.
\newblock


\bibitem[\protect\citeauthoryear{Petitto, Langdon, Stone, Andriola, Kartheiser,
  and Cochran}{Petitto et~al\mbox{.}}{2016}]%
        {petitto2016visual}
\bibfield{author}{\bibinfo{person}{Laura-Ann Petitto}, \bibinfo{person}{Clifton
  Langdon}, \bibinfo{person}{Adam Stone}, \bibinfo{person}{Diana Andriola},
  \bibinfo{person}{Geo Kartheiser}, {and} \bibinfo{person}{Casey Cochran}.}
  \bibinfo{year}{2016}\natexlab{}.
\newblock \showarticletitle{Visual sign phonology: Insights into human reading
  and language from a natural soundless phonology}.
\newblock \bibinfo{journal}{\emph{Wiley Interdisciplinary Reviews: Cognitive
  Science}} \bibinfo{volume}{7}, \bibinfo{number}{6} (\bibinfo{year}{2016}),
  \bibinfo{pages}{366--381}.
\newblock


\bibitem[\protect\citeauthoryear{Petitto and Marentette}{Petitto and
  Marentette}{1991}]%
        {petitto1991babbling}
\bibfield{author}{\bibinfo{person}{Laura~Ann Petitto} {and}
  \bibinfo{person}{Paula~F Marentette}.} \bibinfo{year}{1991}\natexlab{}.
\newblock \showarticletitle{Babbling in the manual mode: Evidence for the
  ontogeny of language}.
\newblock \bibinfo{journal}{\emph{Science}} \bibinfo{volume}{251},
  \bibinfo{number}{5000} (\bibinfo{year}{1991}), \bibinfo{pages}{1493--1496}.
\newblock


\bibitem[\protect\citeauthoryear{Petitto and Neuroimaging}{Petitto and
  Neuroimaging}{[n. d.]}]%
        {petittoimpact}
\bibfield{author}{\bibinfo{person}{Laura-Ann Petitto} {and} \bibinfo{person}{BL
  Neuroimaging}.} \bibinfo{year}{[n. d.]}\natexlab{}.
\newblock \showarticletitle{The Impact of Minimal Language Experience on
  Children During Sensitive Periods of Brain and Early Language Development:
  Myths Debunked and New Policy Implications}.
\newblock  (\bibinfo{year}{[n. d.]}).
\newblock


\bibitem[\protect\citeauthoryear{Pezeshkpour, Marshall, Elliott, and
  Bangham}{Pezeshkpour et~al\mbox{.}}{1999}]%
        {pezeshkpour1999development}
\bibfield{author}{\bibinfo{person}{Farzad Pezeshkpour}, \bibinfo{person}{Ian
  Marshall}, \bibinfo{person}{Ralph Elliott}, {and} \bibinfo{person}{J~Andrew
  Bangham}.} \bibinfo{year}{1999}\natexlab{}.
\newblock \showarticletitle{Development of a legible deaf-signing virtual
  human}. In \bibinfo{booktitle}{\emph{Multimedia Computing and Systems, 1999.
  IEEE International Conference on}}, Vol.~\bibinfo{volume}{1}. IEEE,
  \bibinfo{pages}{333--338}.
\newblock


\bibitem[\protect\citeauthoryear{Rajkumar, Gagliardi, and Sha}{Rajkumar
  et~al\mbox{.}}{1995}]%
        {rajkumar1995real}
\bibfield{author}{\bibinfo{person}{Ragunathan Rajkumar},
  \bibinfo{person}{Michael Gagliardi}, {and} \bibinfo{person}{Lui Sha}.}
  \bibinfo{year}{1995}\natexlab{}.
\newblock \showarticletitle{The real-time publisher/subscriber inter-process
  communication model for distributed real-time systems: design and
  implementation}. In \bibinfo{booktitle}{\emph{Real-Time Technology and
  Applications Symposium, 1995. Proceedings}}. IEEE, \bibinfo{pages}{66--75}.
\newblock


\bibitem[\protect\citeauthoryear{Reilly, McIntire, and Bellugi}{Reilly
  et~al\mbox{.}}{1990}]%
        {reilly1990acquisition}
\bibfield{author}{\bibinfo{person}{Judy~Snitzer Reilly},
  \bibinfo{person}{Marina McIntire}, {and} \bibinfo{person}{Ursula Bellugi}.}
  \bibinfo{year}{1990}\natexlab{}.
\newblock \showarticletitle{The acquisition of conditionals in Language:
  Grammaticized facial expressions}.
\newblock \bibinfo{journal}{\emph{Applied Psycholinguistics}}
  \bibinfo{volume}{11}, \bibinfo{number}{4} (\bibinfo{year}{1990}),
  \bibinfo{pages}{369--392}.
\newblock


\bibitem[\protect\citeauthoryear{Richert, Robb, and Smith}{Richert
  et~al\mbox{.}}{2011}]%
        {richert2011media}
\bibfield{author}{\bibinfo{person}{Rebekah~A Richert},
  \bibinfo{person}{Michael~B Robb}, {and} \bibinfo{person}{Erin~I Smith}.}
  \bibinfo{year}{2011}\natexlab{}.
\newblock \showarticletitle{Media as social partners: The social nature of
  young children's learning from screen media}.
\newblock \bibinfo{journal}{\emph{Child Development}} \bibinfo{volume}{82},
  \bibinfo{number}{1} (\bibinfo{year}{2011}), \bibinfo{pages}{82--95}.
\newblock


\bibitem[\protect\citeauthoryear{Saffran, Senghas, and Trueswell}{Saffran
  et~al\mbox{.}}{2001}]%
        {saffran2001acquisition}
\bibfield{author}{\bibinfo{person}{Jenny~R Saffran}, \bibinfo{person}{Ann
  Senghas}, {and} \bibinfo{person}{John~C Trueswell}.}
  \bibinfo{year}{2001}\natexlab{}.
\newblock \showarticletitle{The acquisition of language by children}.
\newblock \bibinfo{journal}{\emph{Proceedings of the National Academy of
  Sciences}} \bibinfo{volume}{98}, \bibinfo{number}{23} (\bibinfo{year}{2001}),
  \bibinfo{pages}{12874--12875}.
\newblock


\bibitem[\protect\citeauthoryear{Scassellati, Brawer, Tsui, Nasihati~Gilani,
  Malzkuhn, Manini, Stone, Kartheiser, Merla, Shapiro,
  et~al\mbox{.}}{Scassellati et~al\mbox{.}}{2018}]%
        {scassellati2018teaching}
\bibfield{author}{\bibinfo{person}{Brian Scassellati}, \bibinfo{person}{Jake
  Brawer}, \bibinfo{person}{Katherine Tsui}, \bibinfo{person}{Setareh
  Nasihati~Gilani}, \bibinfo{person}{Melissa Malzkuhn},
  \bibinfo{person}{Barbara Manini}, \bibinfo{person}{Adam Stone},
  \bibinfo{person}{Geo Kartheiser}, \bibinfo{person}{Arcangelo Merla},
  \bibinfo{person}{Ari Shapiro}, {et~al\mbox{.}}}
  \bibinfo{year}{2018}\natexlab{}.
\newblock \showarticletitle{Teaching Language to Deaf Infants with a Robot and
  a Virtual Human}. In \bibinfo{booktitle}{\emph{Proceedings of the 2018 CHI
  Conference on Human Factors in Computing Systems}}. ACM,
  \bibinfo{pages}{553}.
\newblock


\bibitem[\protect\citeauthoryear{Schein and Delk~Jr}{Schein and
  Delk~Jr}{1974}]%
        {schein1974deaf}
\bibfield{author}{\bibinfo{person}{Jerome~D Schein} {and}
  \bibinfo{person}{Marcus~T Delk~Jr}.} \bibinfo{year}{1974}\natexlab{}.
\newblock \showarticletitle{The deaf population of the United States.}
\newblock  (\bibinfo{year}{1974}).
\newblock


\bibitem[\protect\citeauthoryear{Schnepp, Wolfe, McDonald, and Toro}{Schnepp
  et~al\mbox{.}}{2013}]%
        {schnepp2013generating}
\bibfield{author}{\bibinfo{person}{Jerry Schnepp}, \bibinfo{person}{Rosalee
  Wolfe}, \bibinfo{person}{John McDonald}, {and} \bibinfo{person}{Jorge Toro}.}
  \bibinfo{year}{2013}\natexlab{}.
\newblock \showarticletitle{Generating co-occurring facial nonmanual signals in
  synthesized American sign language}.
\newblock  (\bibinfo{year}{2013}).
\newblock


\bibitem[\protect\citeauthoryear{Shapiro}{Shapiro}{2011}]%
        {shapiro2011building}
\bibfield{author}{\bibinfo{person}{Ari Shapiro}.}
  \bibinfo{year}{2011}\natexlab{}.
\newblock \showarticletitle{Building a character animation system}. In
  \bibinfo{booktitle}{\emph{INTERNATIONAL Conference on Motion in Games}}.
  Springer, \bibinfo{pages}{98--109}.
\newblock


\bibitem[\protect\citeauthoryear{Starr}{Starr}{2014}]%
        {toshibaRobot}
\bibfield{author}{\bibinfo{person}{Michelle Starr}.}
  \bibinfo{year}{2014}\natexlab{}.
\newblock \bibinfo{title}{Toshiba's new robot can speak in sign language}.
\newblock
  \bibinfo{howpublished}{\url{https://www.cnet.com/news/toshibas-new-robot-can-speak-in-sign-language/}}.
\newblock
\newblock
\shownote{Accessed: 2018-04-25.}


\bibitem[\protect\citeauthoryear{Stone, Petitto, and Bosworth}{Stone
  et~al\mbox{.}}{2018}]%
        {stone2018visual}
\bibfield{author}{\bibinfo{person}{Adam Stone}, \bibinfo{person}{Laura-Ann
  Petitto}, {and} \bibinfo{person}{Rain Bosworth}.}
  \bibinfo{year}{2018}\natexlab{}.
\newblock \showarticletitle{Visual sonority modulates infants' attraction to
  sign language}.
\newblock \bibinfo{journal}{\emph{Language Learning and Development}}
  \bibinfo{volume}{14}, \bibinfo{number}{2} (\bibinfo{year}{2018}),
  \bibinfo{pages}{130--148}.
\newblock


\bibitem[\protect\citeauthoryear{Teena and Manickavasagan}{Teena and
  Manickavasagan}{2014}]%
        {teena2014thermal}
\bibfield{author}{\bibinfo{person}{M Teena} {and} \bibinfo{person}{A
  Manickavasagan}.} \bibinfo{year}{2014}\natexlab{}.
\newblock \showarticletitle{Thermal infrared imaging}.
\newblock In \bibinfo{booktitle}{\emph{Imaging with Electromagnetic Spectrum}}.
  \bibinfo{publisher}{Springer}, \bibinfo{pages}{147--173}.
\newblock


\bibitem[\protect\citeauthoryear{Tobii Eyetracker}{Tobii Eyetracker}{2018}]%
        {tobiiEyetracker2}
Tobii Eyetracker \bibinfo{year}{2018}\natexlab{}.
\newblock \bibinfo{title}{Tobii Pro X3-120}.
\newblock
  \bibinfo{howpublished}{\url{https://www.tobiipro.com/product-listing/tobii-pro-x3-120/}}.
\newblock
\newblock
\shownote{Accessed: 2018-04-25.}


\bibitem[\protect\citeauthoryear{Traum and Larsson}{Traum and Larsson}{2003}]%
        {TraumLarsson03}
\bibfield{author}{\bibinfo{person}{David Traum} {and} \bibinfo{person}{Staffan
  Larsson}.} \bibinfo{year}{2003}\natexlab{}.
\newblock \showarticletitle{The Information State Approach to Dialogue
  Management}.
\newblock In \bibinfo{booktitle}{\emph{Current and New Directions in Discourse
  and Dialogue}}, \bibfield{editor}{\bibinfo{person}{Jan van Kuppevelt} {and}
  \bibinfo{person}{Ronnie Smith}} (Eds.). \bibinfo{publisher}{Kluwer},
  \bibinfo{pages}{325--353}.
\newblock


\bibitem[\protect\citeauthoryear{Uluer, Akal{\i}n, and K{\"o}se}{Uluer
  et~al\mbox{.}}{2015}]%
        {uluer2015new}
\bibfield{author}{\bibinfo{person}{P{\i}nar Uluer}, \bibinfo{person}{Neziha
  Akal{\i}n}, {and} \bibinfo{person}{Hatice K{\"o}se}.}
  \bibinfo{year}{2015}\natexlab{}.
\newblock \showarticletitle{A new robotic platform for sign language tutoring}.
\newblock \bibinfo{journal}{\emph{International Journal of Social Robotics}}
  \bibinfo{volume}{7}, \bibinfo{number}{5} (\bibinfo{year}{2015}),
  \bibinfo{pages}{571--585}.
\newblock


\bibitem[\protect\citeauthoryear{van Zijl and Fourie}{van Zijl and
  Fourie}{2007}]%
        {van2007development}
\bibfield{author}{\bibinfo{person}{Lynette van Zijl} {and}
  \bibinfo{person}{Jaco Fourie}.} \bibinfo{year}{2007}\natexlab{}.
\newblock \showarticletitle{The development of a generic signing avatar}. In
  \bibinfo{booktitle}{\emph{Proceedings of the IASTED International Conference
  on Graphics and Visualization in Engineering, GVE}},
  Vol.~\bibinfo{volume}{7}. \bibinfo{pages}{95--100}.
\newblock


\bibitem[\protect\citeauthoryear{Wilson, Finley, Lawson, Wolford, Eddington,
  and Rabinowitz}{Wilson et~al\mbox{.}}{1991}]%
        {wilson1991better}
\bibfield{author}{\bibinfo{person}{Blake~S Wilson}, \bibinfo{person}{Charles~C
  Finley}, \bibinfo{person}{Dewey~T Lawson}, \bibinfo{person}{Robert~D
  Wolford}, \bibinfo{person}{Donald~K Eddington}, {and}
  \bibinfo{person}{William~M Rabinowitz}.} \bibinfo{year}{1991}\natexlab{}.
\newblock \showarticletitle{Better speech recognition with cochlear implants}.
\newblock \bibinfo{journal}{\emph{Nature}} \bibinfo{volume}{352},
  \bibinfo{number}{6332} (\bibinfo{year}{1991}), \bibinfo{pages}{236--238}.
\newblock


\bibitem[\protect\citeauthoryear{Zhang}{Zhang}{2012}]%
        {zhang2012microsoft}
\bibfield{author}{\bibinfo{person}{Zhengyou Zhang}.}
  \bibinfo{year}{2012}\natexlab{}.
\newblock \showarticletitle{Microsoft kinect sensor and its effect}.
\newblock \bibinfo{journal}{\emph{IEEE multimedia}} \bibinfo{volume}{19},
  \bibinfo{number}{2} (\bibinfo{year}{2012}), \bibinfo{pages}{4--10}.
\newblock


\end{thebibliography}
	
\end{document}